\documentclass[aps,prb,showpacs,twocolumn,eqsecnum,superscriptaddress,floatfix]{revtex4-1}

\usepackage{graphicx}
\usepackage{color}
\usepackage{enumerate}
\usepackage{amsmath}
\usepackage{amssymb}
\usepackage{stmaryrd}
\usepackage{multirow}
\usepackage{float}
\usepackage{subfigure}
\usepackage{longtable,booktabs}
\usepackage{hyperref}
\usepackage{url}%Turn on when output to pdf file
\usepackage{soul} %to cross the text
\def\beq{\begin{equation}}
\def\eeq{\end{equation}}

\begin{document}

\title{Many-body Localization Transition in  Rokhsar-Kivelson-type wave functions}

\author{Xiao Chen} 
\affiliation{Department of Physics and Institute for Condensed Matter Theory, University of Illinois at Urbana-Champaign, 
1110 West Green Street, Urbana, Illinois 61801-3080, USA}
\author{Xiongjie Yu} 
\affiliation{Department of Physics and Institute for Condensed Matter Theory, University of Illinois at Urbana-Champaign, 
1110 West Green Street, Urbana, Illinois 61801-3080, USA}

\author{Gil Young Cho}
\affiliation{Department of Physics and Institute for Condensed Matter Theory, University of Illinois at Urbana-Champaign, 
1110 West Green Street, Urbana, Illinois 61801-3080, USA}
\affiliation{Department of Physics, Korea Advanced Institute of Science and Technology, Daejeon 305-701, Korea}

\author{Bryan K. Clark}
\affiliation{Department of Physics and Institute for Condensed Matter Theory, University of Illinois at Urbana-Champaign, 
1110 West Green Street, Urbana, Illinois 61801-3080, USA}

\author{Eduardo Fradkin}
\affiliation{Department of Physics and Institute for Condensed Matter Theory, University of Illinois at Urbana-Champaign, 
1110 West Green Street, Urbana, Illinois 61801-3080, USA}
\affiliation{Kavli Institute for Theoretical Physics, University of California Santa Barbara, Santa Barbara, California 93106-4030, USA}

\begin{abstract}
We construct a family of 
many-body wave functions to study the many-body localization phase transition. The wave functions have a Rokhsar-Kivelson form, in which the weight for the  configurations are chosen from the 
Gibbs weights of a classical spin glass model, known as the Random Energy Model, multiplied by a random sign structure to represent a highly excited state. These wave functions show a phase transition into an MBL phase.  In addition, we see three regimes of entanglement scaling with subsystem size:  scaling with entanglement corresponding to an infinite temperature thermal phase, constant scaling, and a sub-extensive scaling between these limits.  Near the phase transition point, the fluctuations of the R\'enyi entropies are non-Gaussian.  We find that R\'enyi entropies with different R\'enyi index transition into the MBL phase at different points and have different scaling behavior, suggesting a multifractal behavior. 
\end{abstract}
 
\maketitle
\section{Introduction}

For an isolated quantum many-body system totally decoupled from the environment, the system can act as its own heat bath. 
One way to distinguish systems which reach the thermalized state from those which don't is by studying the entanglement entropy of a subsystem. 
If a subsystem  
is  in thermal equilibrium, the entanglement and thermal entropies must be the same and thus must satisfy a volume law, i.e. , the entropy should scale like $N_A^d$ in $d$ dimensions. This is different from the typical scaling behavior of the entanglement entropy of   
ground states, which under some constraints,  
satisfy an area law scaling.\cite{Bombelli-1986,Srednicki1994,Eisert-2010}

Evidence has recently accumulated for a certain class of interacting systems with quenched disorder which fail to thermalize. These systems go under the name many-body localization\cite{Basko06,Oganesyan07,Pal2010} (MBL) and are the interacting analogue of Anderson insulators.\cite{Anderson-1958}  This failure to thermalize has been attributed to an extensive number of locally conserved charges.\cite{Serbyn2013, Huse2014,Imbrie2014,Ros2015,Chandran2015}  For a review of MBL phases, see the recent review of Ref. \onlinecite{Nandkishore2015}.
In an Anderson insulator, all single-particle eigenstates are exponentially localized  in real space, quantum diffusion is impossible at zero temperature and the system is an insulator on macroscopic scales.\cite{Anderson-1958} In such a state thermalization is not possible (without an external heat bath). It has been suggested that many-body localization results from a similar localization of states in Hilbert space.\cite{Monthus10,Canovi11}

The phenomenon of MBL  is principally (and theoretically) observed in  states deep in the excited state spectrum of a macroscopic system, and hence have an extensive {\em excitation energy} which, following standard (but somewhat inexact) terminology, we will call    `finite energy density' states.  As a function of some tuning parameter (typically disorder), there can be a  phase transition from an ergodic (thermalized) phase to an MBL phase. In contrast to its non-interacting counterpart (the Anderson insulator), an MBL phase transition can occur at finite temperature.\cite{Basko06} We should note that phonons may interfere with the observation of an  MBL phase in solids, but MBL may be physically   realized in  optical lattice systems, see e.g. Ref. \onlinecite{Morong2015}.

In this paper we consider the problem of the MBL phase transition by constructing an ensemble of simple `model' many-body wave functions with a simple structure parameterized by a `disorder' strength, and study the  phase transition as a function of this parameter.  We will show that, in  spite of their simple structure, these model wave function can represent both thermal states and MBL states.    The wave functions that we consider have a structure similar to the Rokhsar-Kivelson (RK)  states\cite{Rokhsar1988} and their generalizations.\cite{Ardonne2004,Fradkin2013}
More specifically, we consider states that are linear superpositions of  quantum states labeled by the classical configurations of a system of $N$ Ising spins, and have the form
\beq
|\Psi_{REM} \rangle = \sum_{\{ \mathcal{C}  \}} W[ \mathcal{C} ] |\mathcal{C} \rangle,
\label{eq:RK}
\eeq
Here the quantum mechanical amplitude $W[\mathcal{C}]$ for a  configuration $\mathcal{C}$ of the Ising spins  is given by the Gibbs weight  of a classical spin glass model known as the Random Energy Model, i.e.
\begin{equation}
W[\mathcal{C}] \propto e^{-\beta E[\mathcal{C}]}
\label{eq:REM-weights}
\end{equation}
where the `energy' $E[\mathcal{C}]$ assigned to the configuration $\mathcal{C}$ is taken to be a random number drawn from a Gaussian distribution. By construction, the amplitudes of these states $W[\mathcal{C}]$ are positive real numbers.
The associated classical spin glass model in infinite space dimension (since each spin is coupled to all the other $N-1$ spins)  is known to have a classical thermodynamic phase transition to a spin glass state.\cite{Derrida1980}   The parameter $\beta$, which in the classical spin glass model is the inverse temperature, but in this work will be used as a parameter of the wave function. Notice that we have not defined a quantum Hamiltonian for which the wave function of Eq.\eqref{eq:RK} is an eigenstate and, hence,  we have not actually defined an  energy for the quantum system. Thus, the `energy' of the Random Energy Model should not be confused with the  energy of the quantum state.

The quantum state $|\Psi_{REM}\rangle$ has the manifestly positive weights shown in Eq.\eqref{eq:REM-weights}. Such a state can be a natural candidate for a ground state of a Hamiltonian but not for a typical excited state whose amplitudes are generally non-positive. To mimic a `typical' state deep in the spectrum of a quantum system, 
 we generalize this construction so that, for a given configuration $C$, the amplitudes for these new states are just a {\em random sign}  
 multiplied by the  amplitude $W(\mathcal{C})$ discussed above (a similar  approach has been used in Ref.\onlinecite{Grover_b2014} and Ref.\onlinecite{ Grover_c2014}.)  We will denote the new wave functions by $|\Psi_{REM+\rm{sign}}[\beta]\rangle$.
Here we will also consider the wave function without random signs denoted by $|\Psi_{REM}[\beta]\rangle$ and compare the physical properties of both types of wave functions.

An advantageous aspect of our approach is that we have more analytical control over this system then is typical in interacting disorder systems. In addition we are  able to perform numerical
calculations with a system size ($\gtrsim 30$) which can only be achieved in other numerical MBL studies with the use of matrix-product states \cite{khemani2015obtaining,yu2015finding}.  These wave functions are also conceptually simple, making them an ideal setting to further our understanding of MBL. 

%\textcolor{red}{
Laumann, Pal and Scardicchio studied numerically the MBL state in the {\em quantum} Random Energy Model and found that the MBL quantum phase transition is distinct from the quantum phase transition to the spin-glass phase.\cite{Laumann2014} Here we will find that  in  the RK wave functions $|\Psi_{REM+\textrm{sign}}\rangle$, although  they are not  actual eigenstates of the quantum REM model, the MBL and  spin-glass transitions are also separate.
%}

In this work we will be focused on three particular aspects of the problem- the transition to the MBL phase, the scaling of the entanglement entropy with subsystem size and the transition from being geometrically delocalized
to localized in the Hilbert space.  
To identify the ergodic and MBL phases, we use the R\'enyi entanglement entropies $S_n$ (where $n$ is the R\'enyi index )
\begin{equation}
S_n=\frac{1}{1-n}\log\mbox{Tr}\rho_A^n
\label{eq:Sn}
\end{equation}
of a subsystem $A$ whose size $N_A < N/2$ is smaller than half of the entire system. 
In the limit $n\to 1$, $S_n$ converges to the von Neumann entropy. In an ergodic system, there is a regime where the R\'enyi entropy obeys a volume law 
which changes linearly as a function of $N_A$.
%EF
%\textcolor{red}{
Bauer and Nayak have argued that, for most states in an MBL phase, their entanglement entropy scales at most as an area law of the subsystem size.\cite{Bauer2013}
We will show below that,
%} 
 in our MBL phase, the R\'enyi entropy is 
sub-extensive as a function of subsystem size and is bounded by a finite constant deep inside the MBL phase (Fig. \ref{fig:schematic}). 
 We take particular note of volume laws at an energy density that corresponds
to infinite temperature (ITV) which scales as $N_A \log 2$.  
The scaling behavior of the entanglement entropy may depend on the subsystem size often showing a crossover from ITV to 
sub-extensive as the subsystem
gets larger.  An important subtlety of our model comes from the lack of correlation length in the REM (inherent in a system at infinite dimension). This makes the relevant parameter to consider in looking for such a crossover
not the absolute size $N_A$ but the ratio $t\equiv N_A/N$ and our results will be quoted as a function of this parameter.

%I don't understand what these sentences mean
%For the wave function satisfying ETH, if we look at the subsystem $A$ which is much smaller than the whole system, 
%the reduced density matrix $\rho_A$ takes a thermal form at the high temperature and thus the eigenvalue of $\rho_A$ has an extensive distribution in the whole Hilbert space of system $A$.  While for the MBL state, if we look at the subsystem A, it will localize around some configurations and the eigenvalue for $\rho_A$ only occupies a small fraction of the Hilbert space.
% The R\'enyi entropy is analogous to the inverse participation ratio defined in Anderson localization, where it is used to characterize the distribution of the amplitude of single-particle wave function in the real space. 

While entanglement entropy %is an `order parameter' 
can be used to  distinguishing MBL from ergodic phases, geometric localization is a measure of compactness of the wave function in Hilbert space.
In an Anderson insulator, the localization is of the single-particle wave function in real space and can be characterized by the inverse participation ratio defined as 
\begin{equation}
Y_n=\int d^dx|\psi(x)|^{2n}
\end{equation}
 where $|\psi(x)|^2$ is the probability distribution of single-particle state in real space.\cite{Thouless1974}
 Generically, $Y_n$ takes the scaling form 
 \begin{equation}
 Y_n\sim N^{-\tau(n)}
 \end{equation}
  where the exponent  $\tau(n)=D_n(n-1)$. For the extensive (delocalized) state, $D_n=d$, while for the localized state $D_n=0$. For the critical single-particle wave function at the mobility edge, $D_n$ has a non-trivial dependence on $n$, and it indicates that the critical wave function has a multifractal nature. \cite{Mirlin2007, Wegner1980, Chamon1996, Castillo1997, Gruzberg2011,Kravtsov2015} For non-interacting systems, this multifractal behavior\cite{Halsey1986, Kadanoff1987}  is also manifest in the R\'enyi entropy for the single-particle critical wave function.\cite{Charkravarty2008, Charkravarty2010, Chen2012}  
  
Here we  will present evidence that multifractal behavior also appears in the entanglement near the phase transition into the  MBL phase by looking at the scaling behavior of the R\'enyi entropies.  Loosely speaking the many-body generalization of this multifractality  measures the degree of localization of states  in the multi-dimensional Hilbert space (in a real-space basis) and not just those of a single particle orbital. Multifractal behavior of weight of a state in a Hilbert space has been studied recently by several authors.\cite{Luitz2014,Torres_Herrera2015} In these studies multifractality is used to characterize the geometry of a state in Hilbert space, i.e. its degree of localization in the Hilbert space. In those works, the Shannon-R\'enyi entropies used to quantify the multifractal behavior of the many-body states is a  measure of the statistical properties of the wave functions as probability distributions and are unrelated to the concept of quantum entanglement.
Multifractality has also been discussed in connection with 
the fidelity of the ground state wave functions  in systems at the infinite-disorder fixed point.\cite{Vasseur2015} So far as we know, multifractal behavior of quantum entanglement  in wave functions close to the MBL transition has not been discussed previously in the literature. This  is one of the main questions that we address in our work.

\subsection{Summary of the main results}

\begin{figure}[hbt]
\centering
\includegraphics[width=.5\textwidth]{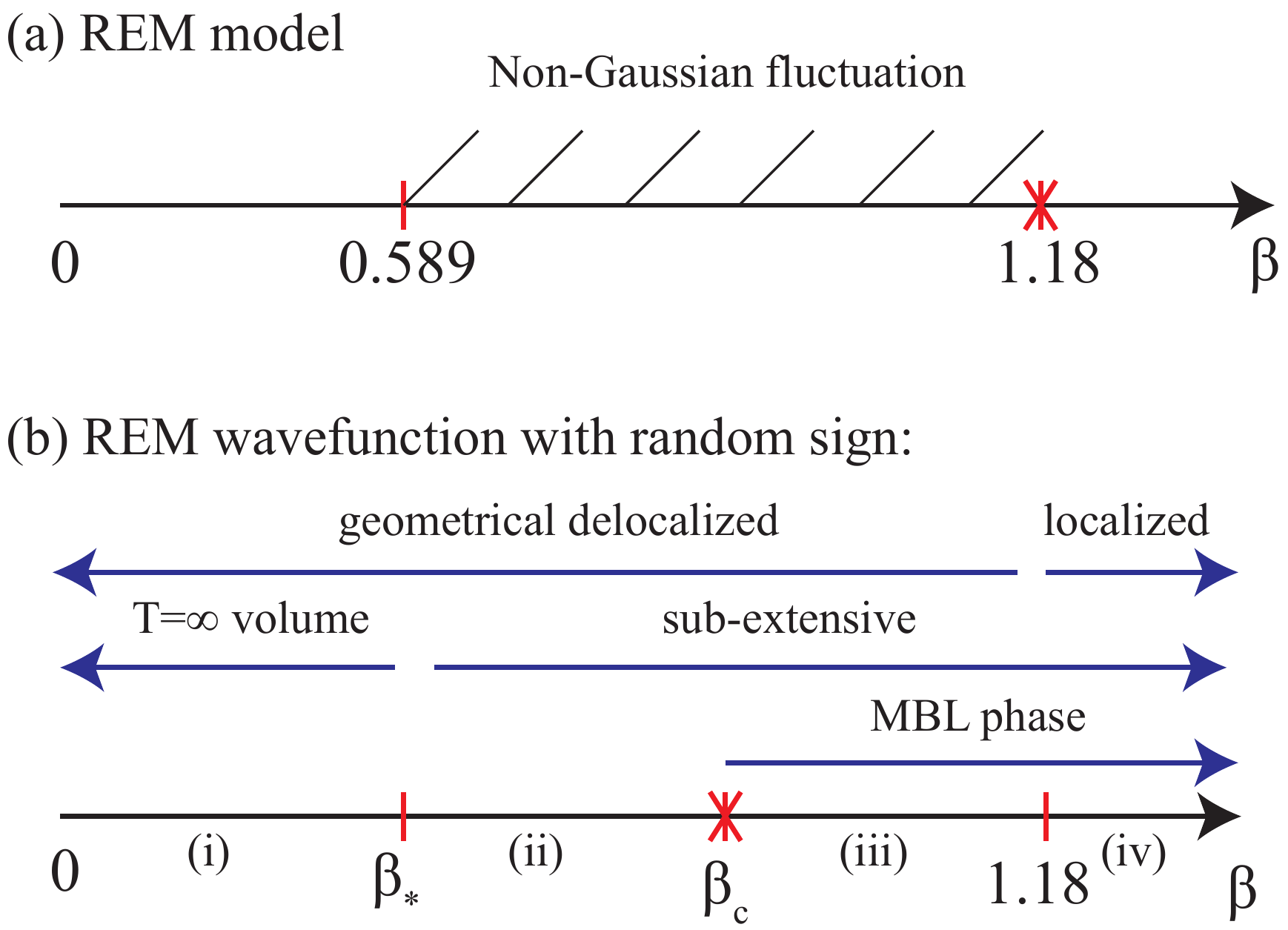}
\caption{(Color online). (a) 
Phase diagram for the classical REM model as a function of $\beta$. The spin glass phase transition occurs at $\beta=\sqrt{2\log2}$. The shaded region between $\sqrt{\log 2/2}$ and $\sqrt{2\log 2}$ has non-Gaussian fluctuation. (b) Schematic phase diagram for the REM wave function in terms of the  entanglement entropy showing different scaling behaviors in four different regimes as a function of $\beta$. In the regime (i) $\langle S_n\rangle$ is equal to $N_A\log2$ for all $N_A$. In the intermediate regime (ii), $\langle S_2\rangle$ is sub-extensive but does not saturate to a constant. Regime (iii) is the MBL phase. In this regime, the wavefunction is not localized in the Hilbert space. (iv) is also the MBL phase with the wavefunction localized in  Hilbert space.
}
\label{fig:schematic}
\end{figure}

In this paper, we consider the MBL transition, the value of $\overline{S(N_A,N)} \equiv \langle S(N_A,N) \rangle $ and the geometric localization of the  wave function $|\Psi_{REM+\textrm{sign}}[\beta]\rangle$.  We show that the wave-function geometrically localizes in Hilbert space at $\beta=1.18$.  We also find three regimes of entanglement scaling for $ \overline{S_2 (N_A;N=\infty)}$: a thermalized regime (at infinite temperature) where the entanglement entropies for a partition show volume law scaling as a function of the subsystem size $N_A$; a regime bounded by constant entanglement entropy; and a regime which is sub-extensive but not constant.  
The transition from sub-extensive to constant happens at, or before, the geometric localization transition.  Notice that the presence of extensive scaling for any $N_A>0$ implies that the finite size scaling of $\overline{S_2(N;N_A)}$ as a function of $N$ at a fixed $N_A$ is also extensive.  Because  $\overline{S_2(\rho_A)}$ is hard to  compute exactly, we analytically calculate  lower and upper bounds for it.  In addition, we identify the MBL transition (with respect to the second Renyi entropy) via a numerical scaling collapse with $\langle S_2 \rangle$ and $\langle \delta S_2 \rangle$ \cite{Kjall2014, Vosk2014} (see Section \ref{sec:finite_size-scaling}).
We find that the the MBL transition is different from the geometric localization transition and therefore the wave-function is still geometrically delocalized in Hilbert space at the MBL transition \cite{Luca2013,Luitz2015} . While the entanglement entropy $\overline{S_2(N_A,N=\infty )}$ scales sub-extensively with $N_A$ at the MBL transition,  the transition to sub-extensive scaling doesn't correspond to the MBL transition.  In addition, the numerical evidence  suggests that the MBL transition (for $S_2$, at $N_A/N=1/2$) happens at a $\beta$ where $\overline{S_2(N;N_A/N=1/2)}$ as a function of $N$ still scales extensively.  While these statements hold for $S_2$, we also consider $S_n$, finding that the MBL transition as well as the analytic bounds for sub-ITV scaling happen for different $n$ at different $\beta$. The former of these appears to scale linearly in $(n-1)/n$. In addition, the scaling exponents identified from scaling collapse are different for different $n$. This is an indication of multifractal behavior. We further study the entanglement spectrum in this regime and observe an entanglement gap between the lower continuous band and the other higher eigenvalues and show that this feature explains transitions which differ for different $n$.
In Fig.\ref{fig:schematic} a and b, we give a succinct and broad summary of the phase diagram (including the phase diagram of the classical REM for comparison).

We also explain the importance of the random sign structure. This effect we compare the behavior of the second R\'enyi entropies $\langle S_2(\rho_A)\rangle$ for $|\Psi_{REM+\textrm{sign}}\rangle$ with $|\Psi_{REM}\rangle$ without random sign. The difference, $\Delta S_2$, is found to decrease monotonically as a function of disorder strength. In the low disorder regimes, $\langle S_2(\rho_A)\rangle$ for $|\Psi_{REM}\rangle$ is a constant and thus their difference $\Delta S_2$ takes the maximal value. At the values of $t \leq 1/3$, the difference disappears before entering into the MBL phase.

%%%%%%%%%%%%%%%%%%%%
%%%%%%%%%%%%%%%%%%%

The rest of the paper is organized as follows. In Section \ref{sec:RK}, we  explain the construction of a Rokhsar-Kivelson-type wave function which assigns to the amplitude of  a quantum many-body state the Gibbs weight  of a classical spin glass model (the Random Energy Model) and introduce a random sign structure into it to access represent typical excited states with a {\em finite excitation energy density}. In Section \ref{sec:REM} we briefly review the classical random energy model (REM) and interpret the classical spin glass phase transition in it. In the Section \ref{sec:random_sign} we  construct a REM wave function with a random sign structure to mimic a highly excited state for a Hamiltonian with disorder. We will analytically compute the R\'enyi entropy to show that there is a thermalized regime and a many-body localized phase. In Section \ref{sec:numerical-results} we  calculate the R\'enyi entropy numerically and find the location of MBL phase transition by finite size scaling. In the Section \ref{sec:random-sign-structure} we  study the R\'enyi entropy for the REM wave function without random sign and demonstrate the importance of the sign structure. In the Section \ref{sec:conclusions} we  summarize our results and conclude that there is a MBL phase transition in the REM wave function with random sign structure.

\section{Rokhsar-Kivelson Model wave functions}
\label{sec:RK}

In the physics of strongly correlated systems there are many examples in which the properties of a new state of matter can be represented by simple `model' wave functions. The best known examples of such model wave functions include the BCS wave function for the ground state of a superconductor and the Laughlin wave function for the fractional quantum Hall fluid. 

\subsection{RK-type wave functions}

In this paper we will give a description of the MBL states using a particularly simple class of model wave functions with a structure similar to the one  proposed by Rokhsar and Kivelson (RK) to capture the physics of the ground states of strongly frustrated quantum antiferromagnets.\cite{Rokhsar1988} In the RK construction the quantum mechanical amplitude of a many-body state is given by a local function of the degrees of freedom, as expressed in the orthonormal basis set $\{ |\mathcal{C} \rangle \} $ where $\mathcal{C}$ is   the `configuration space'. In the RK problem the configuration space  is the set of dimer coverings of a 2D lattice. While in the RK case the dimers are a qualitative representation of spin singlets on each bond of the lattice, a picture of this type has been generalized to many other systems, including Kitaev's Toric Code state,\cite{Kitaev-2003} which represents the topological (or deconfined) phase of a $\mathbb{Z}_2$ gauge theory. 

Since the weights of the RK wave functions are local and positive, they can also be regarded as the Gibbs weights of a related problem in classical statistical mechanics with the same degrees of freedom on the same lattice. Thus if the  basis of orthonormal states is the set $\{ | \mathcal{C} \rangle \}$, i.e. such that $\langle \mathcal{C} | \mathcal{C}' \rangle=\delta_{\mathcal{C},\mathcal{C}'}$, the generalized normalized RK states are 
\begin{equation}
|\Psi \rangle=\frac{1}{\sqrt{\mathcal{Z}}} \sum_{\{ \mathcal{C}\} } e^{-\frac{\beta}{2} E[\mathcal{C}]} |\mathcal{C}\rangle
\label{wave1}
\end{equation}
where 
%For a classical model defined in $d$ spatial dimensions, the partition function can be written as 
\begin{equation}
\mathcal{Z}=\sum_{\{ \mathcal{C} \} }  e^{-\beta E[\mathcal{C}]}
\end{equation}
Here $E[C]$ and $\mathcal{Z}$ are, respectively, the  energy for the classical configuration $\mathcal{C}$ and the partition function  for the associated classical problem, and $\beta$ plays the role of the inverse temperature.
%
%A classical model can be associated to a $d$ dimensional quantum many-body state of the form
%\begin{equation}
%|\Psi_{REM}\rangle=\frac{1}{\sqrt{\mathcal{Z}}}\sum_c e^{-\frac{\beta}{2}E_c}|c\rangle
%\label{wave1}
%\end{equation}
%where the original classical configuration `$c$' maps to an orthonormal quantum state $|c\rangle$ which forms a basis of the Hilbert space 
%and the amplitude for each basis is given by the Boltzmann weight of the classical model. 
%This type of state is a generalized Rokhsar-Kivelson (RK) wave function. 
The original RK wave function is an equal-amplitude superposition of all dimer coverings in $2D$. This wave function can be associated with the partition function for the classical dimer model and is also the ground state of the quantum dimer model at the critical point \cite{Rokhsar1988, Ardonne2004, Fradkin2013} (for 2D bipartite lattices) and a $\mathbb{Z}_2$ topological state\cite{Moessner-2001} (for non-bipartite lattices). States of these type are exact ground states of a special type of quantum Hamiltonians that are the sum of projection operators and are closely related to classical dynamics.\cite{Henley-2004,Castelnovo-2005}

The generalized RK wave function $|\Psi_{REM}\rangle$ inherits many properties from the classical model. For instance, the equal time correlation function of the wave function is the same as the correlation function of the classical model, and the quantum critical point in the wave function corresponds to the classical phase transition at temperature $1/\beta_c$. 
%Other than the original RK state, there are numerous examples for this class of wave function, such as the ground state for Calogero-Sutherland model, \cite{Calogero1969, Sutherland1971} the Laughlin state, the ground state of the quantum Lifshitz model (continuous version of quantum dimer model) \cite{Ardonne2004, Fradkin2013} and the ground state of the Kitaev's toric code model.\cite{Kitaev-2003}

\subsection{Random sign wave function}
\label{sec:random}

We want to consider our family of wave functions as representing finite energy density states with the possibility of supporting both ergodic and MBL phases.  
As we noted above, there exists Hamiltonians constructed by projection operators for which the RK wave functions of the form of Eq.\eqref{wave1} are the exact ground states. However, we should note that its amplitudes are all strictly positive. 
 If the classical Hamiltonian used to generate the RK state only has local interactions, we can show this state must have an area law.  This follows from the fact that one can write a Schmidt decomposition of the state where the number of terms is bounded
by the number of classical interactions which are broken (See Appendix A for detail).  The entanglement properties of these type of states have been discussed in great detail in the case of the quantum dimer model,\cite{Furukawa-2007,Castelnovo-2007,Stephan-2009} of the associated quantum Lifshitz model,\cite{Fradkin2006,Hsu2009,Hsu2010} and of the Toric Code state.\cite{Hamma-2005,Levin2006}

To have the potential for ergodic states, then,  we must either use a non-local classical Hamiltonian or introduce directly a more rich sign structure into the wave function as done, e.g.,  in Refs. \onlinecite{Grover_b2014, Grover_c2014}.  While the classical Hamiltonian we are using for the REM model is non-local, it nonetheless supports an area law at low disorder; in fact, at $\beta=0$, the entanglement entropy is zero over any cut.   Therefore, we introduce a random sign structure in the wave function giving
\begin{equation}
|\Psi_{REM+\rm {sign}}\rangle=\frac{1}{\sqrt{\mathcal{Z}}}\sum_{\{ \mathcal{} C\} }\ s_\mathcal{C} \; e^{-\frac{\beta}{2}E[\mathcal{C}]} |\mathcal{C}\rangle,
\label{wave2}
\end{equation}
where $s_\mathcal{C}$ is a quenched random sign associated to each configuration $\mathcal{C}$. The average ratio between the number of positive and negative signs is one. 
%Although it is not clear if this wave function is the excited state of any known Hamiltonian, the random sign RK state can be considered as the superposition of the highly excited states for RK Hamiltonian which has Eq.\eqref{wave1} as the ground state. 
A similar construction was  discussed in Ref. \onlinecite{Khemani2014}, where they showed that the random sign structure can lead to a thermalized phase. In this paper, we will study the MBL phase transition in a RK-wave function with random sign.

\section{Many body localization phase transition}
\label{sec:MBL}

In this section we will show that the RK wave function with random signs has an ergodic (thermalized) regime and an MBL phase. We begin with a summary of the properties of the classical Random Energy Model whose Boltzmann weights will enter into  the structure our wave function.

\subsection{The Random Energy Model}
\label{sec:REM}

The random energy model (REM) is a simple classical model which  has a phase transition to a spin glass phase.\cite{Derrida1980} This effectively infinite dimensional model has $2^N$ configurations and is the infinite range coupling limit of the Sherrington-Kirkpatrick spin glass model.\cite{Sherrington1975} In the REM model, the energy for each configuration is no longer given by the complicated spin glass Hamiltonian, but rather simply an independent random variable. This random variable has the Gaussian distribution
\begin{equation}
P(E)=(2N\pi)^{-1/2}e^{-\frac{E^2}{2N}}. 
\end{equation}
The number of configurations in the energy  interval  $[N\epsilon, N(\epsilon+\delta)]$, in expectation, is
\begin{equation}
\langle\mathcal{N}(\epsilon,(\epsilon+\delta))\rangle=\int_{\epsilon}^{\epsilon+\delta} dx e^{Ns(x)}
\end{equation}
where 
\begin{equation}
s(x)=\log 2-\frac{x^2}{2}, \qquad \textrm{with}\; x=\frac{E}{N}
\end{equation} 
In the thermodynamic limit, $N\to \infty$, this expectation value takes the asymptotic form
\begin{equation}
\lim_{N \to \infty} \langle\mathcal{N}(\epsilon,(\epsilon+\delta))\rangle=\exp\{N\ \mbox{max}_{x\in[\epsilon,\epsilon+\delta]}s(x)\}
\end{equation}
Notice that $s(x)>0$ only in the interval  $x\in[-\epsilon_*,\epsilon_*]$, where $\epsilon_*=\sqrt{2\log 2}$. This means that for $\epsilon\in [-\epsilon_*,\epsilon_*]$, $\langle\mathcal{N}(\epsilon)\rangle$ is exponentially large, and the fluctuations are very small. For $\epsilon$  outside the interval  $[-\epsilon_*,\epsilon_*]$, $\mathcal{N}(\epsilon)$ is exponentially small.

 The partition function for this model is simply given by
\begin{equation}
\mathcal{Z}=\sum_{i=1}^{2^N}e^{-\beta E_i}=\int dE \; \mathcal{N}(E)\; e^{-\beta E}=\int dx \; e^{N\phi(x)}. 
\label{Z_rem}
\end{equation}
where $\phi(x)$ equals to
\begin{equation}
\phi(x)=\log 2-\frac{x^2}{2}-\beta x, 
\end{equation}

Similar to the calculation for $\langle\mathcal{N}(\epsilon)\rangle$, we can also use the saddle point approximation to calculate the partition function to obtain
\begin{equation}
\mathcal{Z}= \exp\{ N\ \mbox{max}[\phi(x)]\}
\end{equation}
 which is the exact result in the thermodynamic limit, $N \to \infty$. By computing $\phi_{{\rm max}}=\mbox{max}[\phi(x)]$, it is easy to show that the free energy density equals to
\begin{eqnarray}
\nonumber f(\beta)&=&-\frac{\log \mathcal{Z}}{\beta N}=-\frac{\phi_{\rm {max}}}{\beta}\\
&=&\begin{cases}-\frac{\beta}{2}-\frac{\log 2}{\beta} & \beta<\beta_{sg}\\-(2\log 2)^{1/2} & \beta\geq \beta_{sg}\end{cases}
\label{free_energy}
\end{eqnarray}
At $\beta_{sg}=\sqrt{2\log2}$, there is a discontinuity in the second derivative of the free energy density, showing that there is a phase transition at this point, the spin glass transition.

This phase transition can be further studied by computing the inverse participation ratio (IPR), \cite{Derrida1981, Mezard2009}
\begin{equation}
Y_n(\beta)\equiv \sum_i^{2^N}p_i^n=\frac{\sum_i^{2^N} e^{-n\beta E_i}}{(\sum_i^{2^N} e^{-\beta E_i})^n}. 
\label{parti_ratio}
\end{equation}
This quantity measures how many configurations effectively contribute to the partition function and measurable quantities. The expectation value for $Y_2$ equals to \cite{Mezard2009}
\begin{eqnarray}
\langle Y_2(\beta)\rangle=\begin{cases}0  & \beta<\beta_{sg}\\1-\frac{\beta_{sg}}{\beta} & \beta\geq \beta_{sg}\end{cases}
\end{eqnarray}
At  low temperatures, $\beta>\beta_{sg}$, the participation ratio takes a finite value. This means that the system is completely frozen to $O(1)$ number of configurations and is in a non-ergodic phase. At high temperature, $\beta<\beta_{sg}$, all configurations will contribute to the thermodynamic properties of the system. For instance, in the high temperature limit $\beta \to 0$, the Boltzmann's measure becomes uniform and the second  participation ratio becomes $Y_2(\beta\to 0)=2^{-N}$. When $N \to \infty$, $Y_2\to 0$. In general, when $\beta<\beta_{sg}$, $Y_n$ scales with the system size 
\begin{equation}
Y_n\sim \mathcal{D}^{-\tau(n)} , \qquad \textrm{with}\; \mathcal{D}=2^N
\label{eq:multifractal}
\end{equation}
From the results on $\log \mathcal{Z}(\beta)$, we can compute the exponent $\tau(n)$ (for $n>1$)
\begin{equation}
\tau(n)=\begin{cases} (n-1)(1-\gamma n), & 0\leq\gamma <\frac{1}{n^2}\\
n(1-\sqrt{\gamma})^2, & \frac{1}{n^2}<\gamma<1\\
0, & \gamma>1\end{cases}
\label{multi_spec}
\end{equation}
where $\gamma=\frac{\beta^2}{2\log 2}$. We will use $\tau(n)$ to give a upper bound for the R\'enyi entropy of the wave function $|\Psi_{REM+\textrm{sign}}\rangle$ later.

Although the REM model is a simple toy model, it has a spin glass phase transition, i.e., it undergoes a localization transition. It also shows a rich structure in the fluctuation of the free energy. According to the results of Ref. \onlinecite{Bovier2002}, for the ergodic phase, which occurs for $\beta<\beta_{sg}$, there are two regimes with different fluctuation behavior of the free energy. When $\beta<\sqrt{\log 2/2}$, the fluctuations of the free energy are Gaussian, and satisfy the central limit theorem. When $\beta_{sg}>\beta>\sqrt{\log 2/2}$, there are non-Gaussian fluctuations of the free energy driven by the Poisson process of the extreme values of the random energies. This regime does not satisfy the central limit theorem. The resulting phase diagram of classical REM model is shown in Fig.\ref{fig:schematic} (a).

\subsection{Random sign REM wave function}
\label{sec:random_sign}

We can now construct a quantum state following the  procedure  of Eq.\eqref{wave2},
\begin{equation}
|\Psi_{REM}\rangle=\frac{1}{\sqrt{\mathcal{Z}}}\sum_{\{ \mathcal{C} \} } e^{-\frac{\beta}{2}E[\mathcal{C}]} |\mathcal{C}\rangle
\label{wave_rem}
\end{equation}
whose amplitudes are the Boltzmann weights of the classical REM.
By further introducing the random sign structure, the wave function takes the form
\begin{equation}
|\Psi_{REM+\textrm{sign}}\rangle=\frac{1}{\sqrt{\mathcal{Z}}}\sum_{\{ \mathcal{C} \} } s_\mathcal{C} e^{-\frac{\beta}{2}E[\mathcal{C}]}|\mathcal{C}\rangle
\label{wave_rem_sign}
\end{equation}
Here both $s_\mathcal{C}$ and $E[\mathcal{C}]$ are random variables and are independent of each other. The random sign $s_\mathcal{C}$ takes the values $\pm1$  with equal probability over the entire  Hilbert space of $2^N$ spin configurations.

Before we do any calculations, we can first estimate the scaling behavior of the R\'enyi entropy for this quantum state in the extreme limits. In this wave function, $\beta$ is a tuning parameter and has the physical meaning of the disorder strength. The random sign structure is used to represent a highly excited quantum state. When there is no disorder, i.e., $\beta=0$, the amplitude for every configuration is the same. At this point, the R\'enyi entropy for the subsystem $A$ with different R\'enyi index is equal to the thermal entropy at infinite temperature. Thus this wave function is thermalized, and the entanglement entropies obey a volume law.\cite{Grover_b2014}  As $\beta$ increases, the disorder becomes stronger and the entanglement entropy becomes smaller. The wave function is eventually localized to a small fraction of the configurations as $\beta>\beta_{sg}$. As we already showed in the previous section, since the number of these configurations is only $O(1)$, in this regime the R\'enyi entropy is bounded by a finite constant.

\subsubsection{Nomenclature and thermodynamic limit for scaling}

The entanglement entropy $\langle S_n \rangle$ at fixed $\beta$ depends on both the subsystem size $N_A$ as well as the total system size $N$ and there can be separate functional forms for the scaling of  $\langle S_n \rangle$ with respect to either of these parameters. While in many physical systems, these scalings coincide, this is not the case in our model and so it is important to be clear about the distinction.  Unless otherwise specified, our discussion will always focus on scaling with $N_A$ at fixed system size $N$.   

For the thermodynamic limit considered in this work, we let both $N_A$ and $N$ go to infinity but their ratio $t=N_A/N$ to be a finite value. The changes in entanglement scaling we identify then happen at particular values of $t $ for $0 \leq t \leq 1/2$ for different values of $\beta$. The MBL phase in our terminology is identified with scaling collapse.

\subsubsection{Analytic Bounds}

We now analytically compute  bounds to the R\'enyi entropy. Notice that for a disordered system, we need to take a quenched  ensemble average of the R\'enyi entropy,
\begin{equation}
\langle S_{n}(\rho_A)\rangle=\frac{1}{1-n}\langle \log\mbox{Tr}\rho_A^n\rangle
\label{eq:quenched}
\end{equation}
In this quenched average the disorder is frozen and does not evolve with time. We mainly focus here on the second R\'enyi entropy, but the results are easily extended to the other R\'enyi entropies. 

We begin by noting that the quenched average in Eq.\eqref{eq:quenched}  is redundant if, in the thermodynamic limit, the system is self-averaging.\cite{Buffet1993} However, this will not always be the case.  For brevity, we  sometimes will denote $\langle S_n(\rho_A)\rangle\equiv\langle S_n\rangle$.

Let us define a bipartition our system of $N$ spins into two subsets (or regions), $A$ and $B$.
The reduced density matrix for region $A$ is
\begin{eqnarray}
\nonumber \rho_{a,a^{\prime}}^A&=&\frac{1}{\mathcal{Z}}\left(\sum_bs_{a,b}s_{a^{\prime},b}e^{-\frac{\beta}{2}(E_{a,b}+E_{a^{\prime},b})}\right)|\mathcal{C}_a\rangle\otimes\langle \mathcal{C}_{a^{\prime}}|\\
&=&\frac{\widetilde{\rho}^A_{a,a^{\prime}}}{\mathcal{Z}}
\label{rho_A}
\end{eqnarray}
where $\widetilde{\rho}^A_{a,a^{\prime}}$ is the unnormalized reduced density matrix.

For the above reduced density matrix, $\langle S_n(\rho_A)\rangle$ equals to
\begin{equation}
\langle S_n(\rho_A)\rangle=\frac{1}{1-n}\left(\langle\log\mbox{Tr}\widetilde{\rho}_A^n\rangle-n\langle\log\mathcal{Z}(\beta)\rangle\right)
\end{equation}
The second term $\langle\log\mathcal{Z}(\beta)\rangle$ can be calculated by the saddle point approximation and the result is already shown in Eq.\eqref{free_energy}. However, the first term $\langle\log\mbox{Tr}\widetilde{\rho}_A^n\rangle$ is hard to obtain analytically. Instead of calculating it directly, we compute a lower bound and upper bound for it. 

%\begin{itemize}\item {\bf
\paragraph*{Lower Bound}: Consider the {\em annealed average}
\begin{equation}
S_n(\langle \widetilde{\rho}_A\rangle)=\frac{1}{1-n}\left(\log\langle\mbox{Tr}\widetilde{\rho}_A^n\rangle-n\langle\log\mathcal{Z}(\beta)\rangle\right)
\label{anneal}
\end{equation}
which is much easier to compute. By using Jensen's inequality,\cite{Reed-Simon} when $n>1$, it is straightforward to see that  the annealed average of the R\'enyi entropy $S_n(\langle \widetilde{\rho}_A\rangle)$ provides a lower bound for the quenched average $\langle S_n(\rho_A)\rangle$, i.e.
\begin{equation}
\langle S_n(\rho_A)\rangle \geq S_n(\langle \widetilde{\rho}_A\rangle)
\end{equation}

To obtain  the annealed average $S_2(\langle \widetilde{\rho}_A\rangle)$, we need to calculate $\langle\mbox{Tr}\widetilde{\rho}_A^2\rangle$. By means of simple manipulations we find
\begin{align}
\langle\mbox{Tr}\widetilde{\rho}_A^2\rangle=&\sum_{a,a^{\prime}}\langle(\widetilde{\rho}^A_{a,a^{\prime}})^2\rangle\nonumber\\
=& \sum_{a}\langle\left(\widetilde{\rho}_{a,a}^A\right)^2\rangle+ \sum_{a\neq a^{\prime}}\langle\left(\widetilde{\rho}^A_{a,a^{\prime}}\right)^2\rangle \nonumber \\
=&\sum_a \langle\Big(\sum_be^{-\beta E_{a,b}}\Big)^2\rangle\nonumber\\
&+\sum_{a\neq a^{\prime}}\langle\Big(\sum_b s_{a,b}s_{a^{\prime},b}e^{-\frac{\beta}{2}(E_{a,b}+E_{a^{\prime},b})}\Big)^2\rangle \nonumber \\
=&2^Ne^{2N\beta^2}+(2^{N_B}-1)2^{N}e^{N\beta^2}+(2^{N_A}-1)2^{N}e^{N\beta^2}
\label{random_sign}
\end{align}
where the last step is derived by using that for a Gaussian distribution $\langle e^{-\alpha E}\rangle=e^{\alpha^2N/2}$. 

\begin{figure}[hbt]
\centering
\includegraphics[scale=.32]{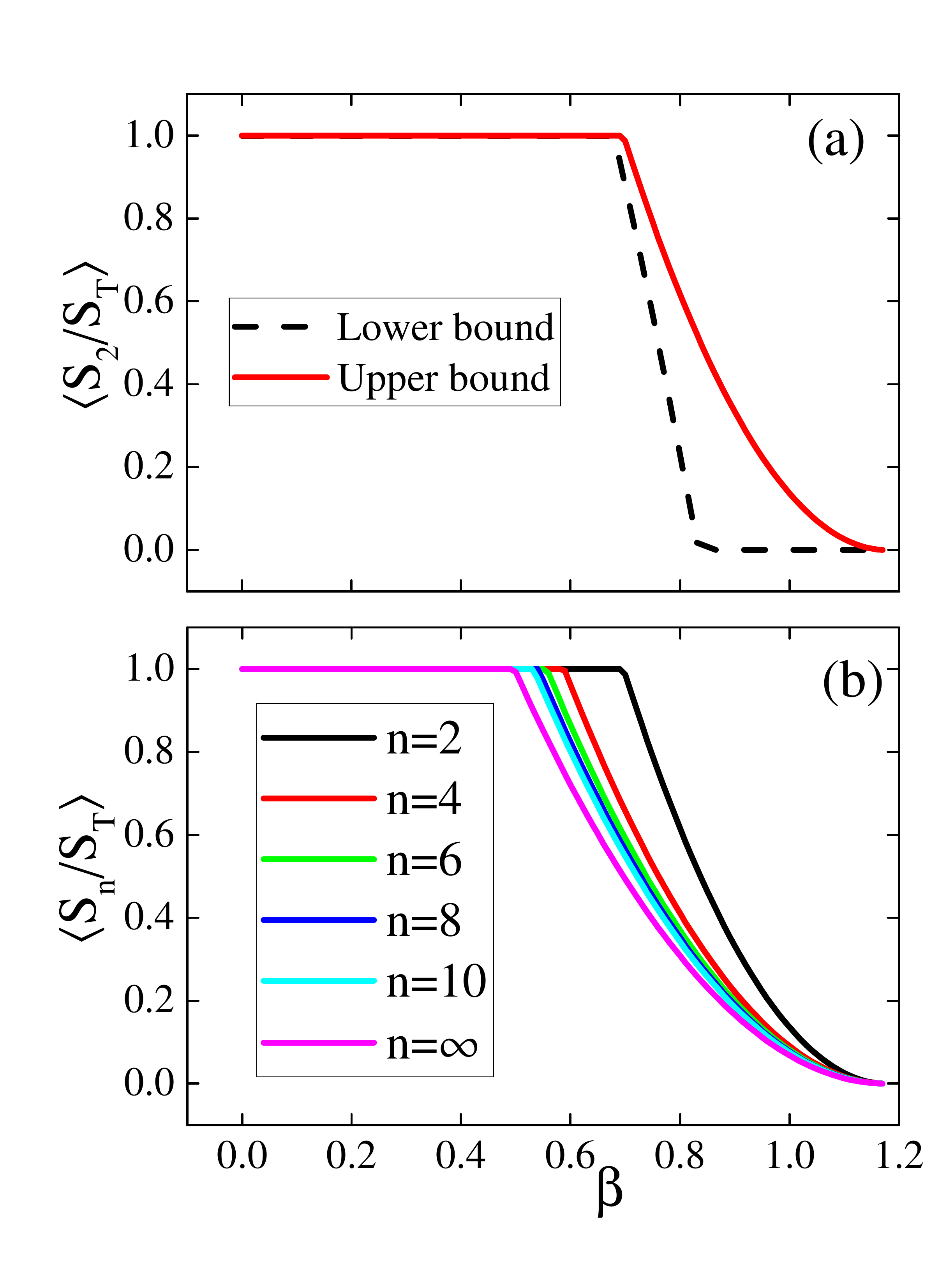}
\caption{(Color online) (a) The lower and upper bound for $\langle S_2\rangle$. The black dashed curve is for the lower bound and the red solid curve is for the upper bound. The system size $N=300$ and the ratio $t=1/3$. (b) The upper bound for $\langle S_n\rangle$. The setup is the same as (a).}
\label{S_n_bound}
\end{figure}

In the thermodynamic limit, when $N_A<N$, the annealed average of the second R\'enyi entropy becomes
\begin{equation}
S_2(\langle \widetilde{\rho}_A\rangle)=\begin{cases} N_A\log2, & \beta\leq\beta_1\\
N(\log2-\beta^2), & \beta>\beta_1
\end{cases}
\label{lower_b}
\end{equation}
where $\beta_1=\sqrt{(1-t)\log2}$ and $t$ is the ratio $N_A/N$. The result is plotted in Fig.\ref{S_n_bound} (a), where the black dashed curve is the lower bound for $\langle S_2\rangle/S_T$ at $t=1/3$. $S_T=N_A\log2$ is the thermal entropy for subsystem $N_A$ at infinite temperature.  Notice that when $\beta\leq\beta_1$, since the lower bound $S_2(\langle \widetilde{\rho}_A\rangle)=S_T$, $\langle S_2(\rho_A)\rangle$ must be equal to $N_A\log2$. When $\beta>\sqrt{\log2}$, $S_2(\langle \widetilde{\rho}_A\rangle)<0$, this lower bound is replaced by zero and is not useful anymore.

%\item {\bf
\paragraph*{Upper Bound}:  It is easy to see that $\mbox{Tr}\rho_A^2$ can be bounded from below by the second participation ratio of the classical REM. Indeed,
\begin{eqnarray}
 \mbox{Tr} \rho_A^2 > \frac{\sum_a (\sum_be^{-\beta E_{a,b}})^2}{\mathcal{Z}(\beta)^2}>\frac{\mathcal{Z}(2\beta)}{\mathcal{Z}(\beta)^2}=Y_2(\beta)
\end{eqnarray}
Thus, an upper bound for the quenched average $\langle S_2(\rho_A)\rangle$ is given by 
\begin{equation}
\langle S_2(\rho_A)\rangle < \tau_2(\beta)N\log2
\end{equation}
where $\tau_2(\beta)$ is defined in Eq.\eqref{multi_spec}.  This upper bound puts a constraint on $\langle S_2(\rho_A)\rangle$, showing that when 
\begin{equation}
\beta>\beta_{ub}(n=2)=(1-\sqrt{\frac{t}{2}})\sqrt{2\log 2}
\label{upper_b}
\end{equation}
then
\begin{equation}
\langle S_2(\rho_A)\rangle < N_A\log2
\end{equation}
 which is the thermal entropy at infinite temperature. The red solid curve in Fig.\ref{S_n_bound} (a) is the upper bound for $\langle S_2\rangle/S_T$ at $t=1/3$. 

%%%%%%%%%%%%%%%%%%%%
%%%%%%%%%%%%%%%%%%%%

%%%%%%%%%%%%%%%%%%%%
\begin{figure}[hbt]
\centering
\includegraphics[scale=.32]{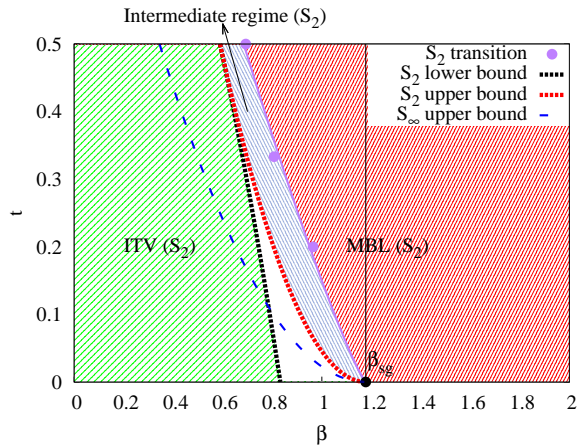}
\caption{(Color online) A summary of our knowledge of the phase diagram of $\Psi_{REM+\textrm{sign}}$, based on $\langle S_n\rangle$.  The dotted black line indicates a bound to the left of which $\langle S_2\rangle$ analytically follows the $T=\infty$ volume law (ITV).  The dotted red line indicates a bound to the right of which $\langle S_2\rangle$ analytically is strictly below the $T=\infty$ volume law.  $\beta_{sg}$ indicates analytically the transition to a localized Hilbert space for any $N_A/N$ and a guaranteed constant for all $\langle S_n\rangle$. The solid purple dots indicate the numerically computed transition points for $\langle S_2\rangle$ (the purple line is a fit for the eye). In comparison, the blue dashed line indicates a bound to the right of which $\langle S_\infty\rangle$ analytically is strictly less then ITV, and suggests multifractality.}
\label{summary}
\end{figure}
%%%%%%%%%%%%%%%%%%%%
Using a similar approach, we can prove that when $n$ is an even number, $\mbox{Tr}\rho_A^n\geq Y_n(\beta)$, and the quenched averaged $n$th R\'enyi entropy satisfies
\begin{eqnarray}
\langle S_n(\rho_A)\rangle < \frac{\tau_n(\beta)N\log2}{n-1}
\label{S_n_upper}
\end{eqnarray}
 This upper bound indicates that when 
\begin{equation}
\beta>\beta_{ub}(n)=(1-\sqrt{\frac{n-1}{n}t})\sqrt{2\log 2}
\label{S_n_upper_b}
\end{equation}
we obtain the bound
\begin{equation}
\langle S_n(\rho_A)\rangle < N_A\log2
\end{equation}
 which, again, is the thermal entropy at the infinite temperature. The result for the upper bound is shown in Fig.\ref{S_n_bound} (b). Notice that when $\beta\leq\beta_{ub}(n=\infty)$, the different curves in Fig.\ref{S_n_bound} (b) are overlapping with each other. When $\beta_{ub}(n=\infty)<\beta<\beta_{sg}$, the upper bounds for $\langle S_n\rangle$ show different scaling behavior.
Combining the upper and lower bounds for different $n$ leads to a regime where the $n=2$ R\'enyi entropy satisfies a volume law and behaves like a thermal entropy at infinite temperature,  but for $n>2$ is strictly below this bound.

%\end{itemize}
\paragraph{Implications from the bounds}: From the above results, we  find an upper  and lower bound for the R\'enyi entropy.  Here we consider $\langle S_2\rangle$ in detail, which thus has the following behaviors
\begin{equation}
\begin{cases}\langle S_2\rangle=S_2(\langle \widetilde{\rho}_A\rangle)=N_A\log2, & \beta\leq\beta_1\\
N(\log 2-\beta^2)<\langle S_2\rangle<N_A\log2, &\beta_2<\beta<\beta_{sg}\\
\langle S_2\rangle<-\log(1-\beta_{sg}/\beta), &\beta>\beta_{sg}
\end{cases}
\label{s_2}
\end{equation}
where $\beta_2=\beta_{ub}(n=2)=(1-\sqrt{t/2})\sqrt{2\log 2}$.

For $\langle S_2\rangle$, there is a regime where $\langle S_2\rangle$ satisfies the volume law and equals the thermal entropy at $T=\infty$. 
When $\beta>\beta_{sg}$, $\langle S_2\rangle$ is bounded by a finite constant.  A phase transition into the MBL phase is expected to be between $\beta_1\leq \beta\leq\beta_{sg}$ .  To find the location of the MBL phase transition, we will use scaling collapse in  Section \ref{sec:numerical-results}.

\subsection{Localization properties of the wave function}
\label{sec:localization}

We end this Section with a discussion of the statistical properties of the states $|\Psi_{REM}\rangle $ and $|\Psi_{REM+\textrm{sign}}\rangle$. Given these states, we can define the amplitudes, i.e. their overlap with an eigenstate of the spins  $|\mathcal{C}\rangle$. For each state, the square of the amplitude  defines a probability distribution for the configuration $\mathcal{C}$ to occur in the state (and hence the random sign does not affect the probability distribution). The resulting probability distribution is thus the same for both states, and it is given by the probability of the configuration $\mathcal{C}$ in the classical REM,
\begin{equation}
P[\mathcal{C}]=|\langle \mathcal{C}| \Psi_{REM}\rangle|^2=\frac{1}{\mathcal{Z}} e^{-\beta E[\mathcal{C}]}
\label{sec:probability}
\end{equation}
Given this probability distribution, we can compute its Shannon-R\'enyi entropies, which one can immediately see, c.f. Eq.\eqref{parti_ratio}, to be the same as the IPR of the classical REM.
 Hence, the localization of wave function in the Hilbert space can be characterized by the IPR defined in the configuration space. For REM wave function, 
\begin{equation}
Y_n=\sum_{\{ \mathcal{C}\} } |\langle\Psi|\mathcal{C}\rangle|^{2n}=\frac{\sum_{i=1}^{2^N} e^{-n\beta E_i}}{\left(\sum_{i=1}^{2^N} e^{-\beta E_i}\right)^n}
\end{equation}
This is the same as the IPR for the classical REM model defined in Eq.\eqref{parti_ratio}. Since $\tau(n)$, defined in Eq.\eqref{eq:multifractal},  is the multifractal spectrum, given explicitly in Eq.\eqref{multi_spec}, in the regime $0<\beta<\beta_{sg}$, the wave function itself has multifractal behavior. The multifractality of the wave function indicates the pre-freezing behavior before entering into the MBL phase. Similar phenomenon was also observed in Refs.\onlinecite{Luca2013,Luitz2015}, where they found that at the MBL phase transition point, the whole wave function is still  delocalized in the configuration space. 

We can now use the results of the inverse participation ratios of the classical REM summarized in section \ref{sec:REM} to draw conclusions on the degree of localization in the $2^N$-dimensional Hilbert space of the $|\Psi_{REM}\rangle $ and $|\Psi_{REM+\textrm{sign}}\rangle$ wave functions. From the results of Section \ref{sec:REM} we find that for $\beta > \beta_{sg}=\sqrt{2 \log 2}$ all the inverse participation ratios are finite as $N \to \infty$ and, hence, that the Shannon-R\'enyi entropies for the wave functions are finite (and are not extensive). Thus, in this regime   these wave functions are (exponentially) localized. On the other hand, for $\beta <\beta_{sg}$, the IPRs of the REM vanish exponentially fast as $N \to \infty$, and so do the Shannon-R\'enyi entropies of the wave functions. In this regime the wave functions are not localized. In this regime the multifractal nature of these wave functions is manifest in the size dependence of their Shannon-R\'enyi entropies. We should emphasize that, beyond setting  the bounds that we have discussed earlier in this section, the knowledge the behavior of the Shannon-R\'enyi entropies alone yields no information on the scaling of quantum entanglement, which is our main interest.

\section{Numerical results on MBL phase transition}
\label{sec:numerical-results}

In the previous section, we were able to analytically establish the existence of a regime in which the quenched averaged second R\'enyi entropy $\langle S_2\rangle$  is strictly less than ITV scaling, and happens strictly before the localization transition.  Nothing prevents this regime from being one in which $\langle S_2(N_A) \rangle$ still scales linearly but at finite $T$; in addition, it doesn't separate the MBL transition  from the localization transition. %and so unfortunately, this
%does not resolve whether the system becomes sub-extensive prior to the localization transition.
 To establish this, we numerically identify  the transition point via scaling collapse. We explicitly construct different disorder samples from the random sign REM wave function and calculate the quenched average $\langle S_n\rangle$ with at least 1000 disorder realizations. First, we focus on the $n=2$ case, and then discuss the other values of $n$.
As a sanity check, we verify that our numerical results are consistent with the analytical results above at large and small $\beta$.

\begin{figure}[hbt]
\centering
\includegraphics[scale=.32]{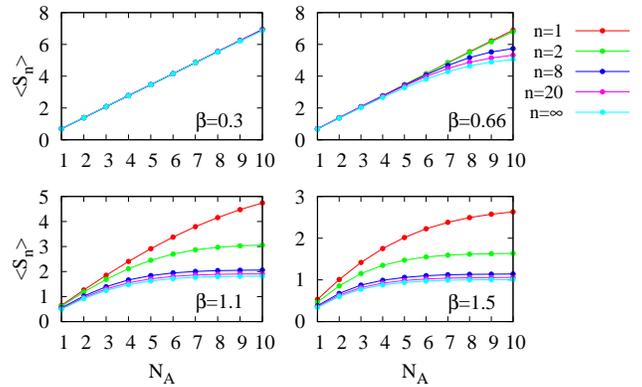}
\caption{(Color online) The numerical results for $\langle S_n\rangle$ with $n$ from $1$ to $\infty$ at different values of $\beta$. The total system size is $N=30$. Each point in the plot is averaged over 2000 disorder realizations. At $\beta=0.3$, $S_n$ shows ITV behavior for all $n$. At $\beta=0.66,1.1$, $\langle S_n \rangle$ deviates from the ITV with increasing $\beta$, but the speed at which it moves away depends on its R\'enyi index. At $\beta=1.5$, $\langle S_n \rangle$ is a constant for all $n$.}
\label{B03}
\label{B15}
\label{B11}
\label{B066}
\end{figure}

The following is a summary of our  numerical results.
\paragraph*{At small $\beta$}:  Analytically we anticipate ITV for $n=2$. In Fig. \ref{B03}, we concretely consider $\beta=0.3$ at $N=30$ with 2000 disorder realizations.  We see that it corresponds to the expected
$\langle S_2 \rangle = N_A \log 2$.

As $N_A \log 2$ is the maximal entanglement entropy for any given realization, for a $T=\infty$  volume law to hold, essentially  all but a measure zero fraction of configurations must have this entropy.  From the  inset of Fig.\ref{S2_scaled}b, one can see that the standard deviation of $\langle S_2\rangle$ at $\beta=0.3$ is zero showing this is indeed the case. Similar behavior can be observed for $\langle S_n\rangle$ with other R\'enyi indices, where $\langle S_n\rangle=N_A\log2$ and $\langle\delta S_n\rangle$ is close to zero as $\beta\leq\beta_*$. 

\paragraph*{At large $\beta>\beta_{sg}$}:  We  find analytically that $\langle S_n\rangle$ with $n\geq 2$ is bounded by a finite constant and, hence, in this range the state is in the MBL phase.  Note that a constant for small $n$ gives a bound for the entanglement entropy for all larger $m$ as $S_n>S_m$ if $m>n$.  In  Fig.\ref{B03} the numerical results for $\beta=1.5$ are presented, and show that when $N_A$ increases, $\langle S_n\rangle$  saturates to some constant value.  Notice also from the  inset of Fig.\ref{S2_scaled}b, that as we move deeper into the localized phase, the standard deviation is monotonically decreasing.

\begin{figure}[hbt]
\centering
\includegraphics[scale=0.9]{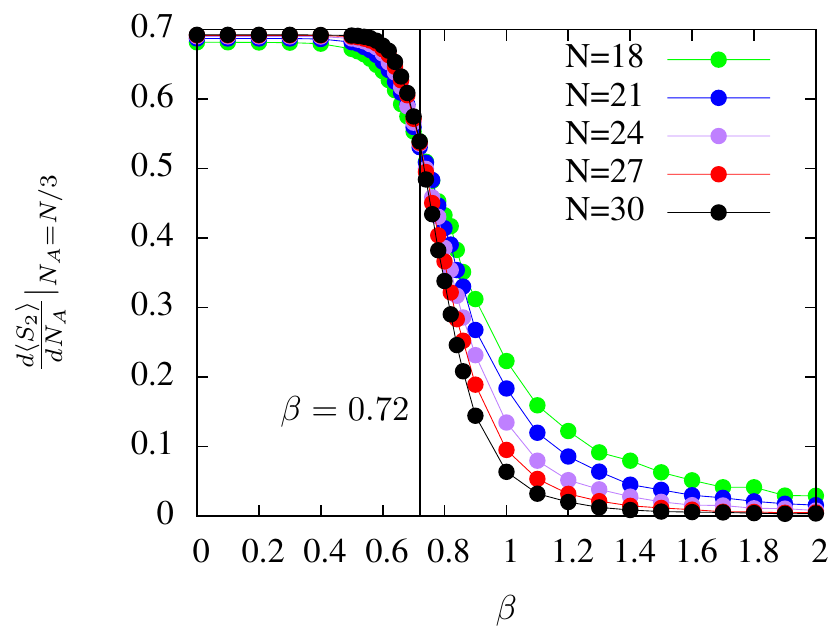}
\caption{(Color online) $\frac{d\langle S_2 \rangle}{dN_A}$ vs $\beta$ graph at $t=1/3$, where the slopes are obtained from finite differences. As $\beta$ increases, the slopes drop from $\log2$ towards 0 for all 5 different system sizes.}
\label{S2_slope}
\end{figure}

The change from ITV to constant in the entanglement entropy can be seen in Fig.\ref{S2_slope}, which shows $d\langle S_2\rangle/d N_A$ as a function of $\beta$. As system sizes increases, the slope quickly approaches $\log2$ for $\beta < 0.5$, and approaches $0$ for $\beta>1.2$. The slopes for all 5 values of $N$ start to drop around $\beta=0.6$, which is less than the $\beta_2=(1-\sqrt{1/6})\sqrt{2 \log2} \approx 0.6967$ in Eq.\eqref{s_2} for subsystem ratio $t=1/3$ and is consistent with the analytical result. 

%%%%%%%%%%%%%%%%%%%%
%%%%%%%%%%%%%%%%%%%%

\subsection{Finite-size scaling}
\label{sec:finite_size-scaling}

\begin{figure}[hbt]
\centering
\subfigure[]{\includegraphics[scale=.3]{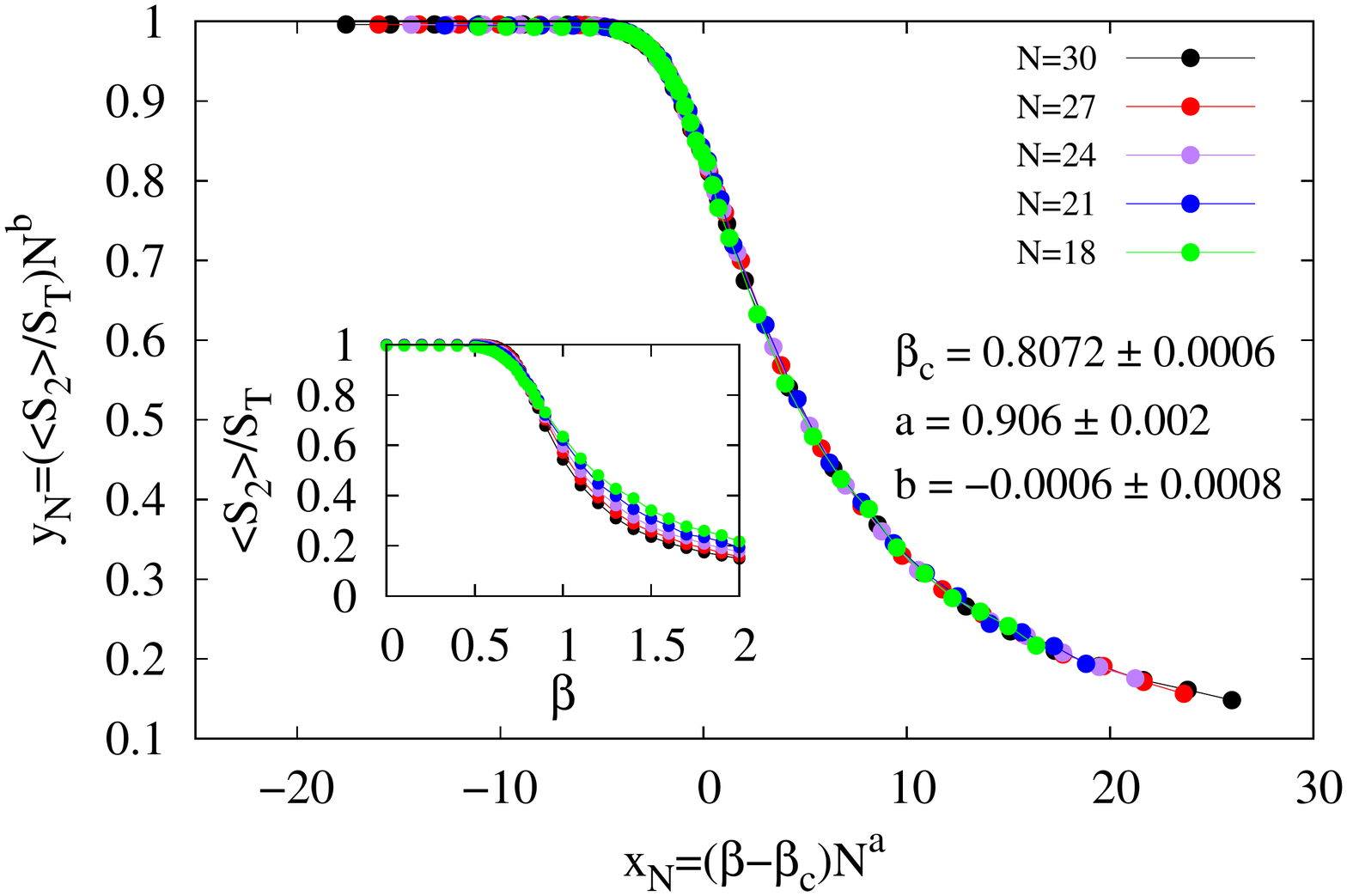}}
\subfigure[]{\includegraphics[scale=.3]{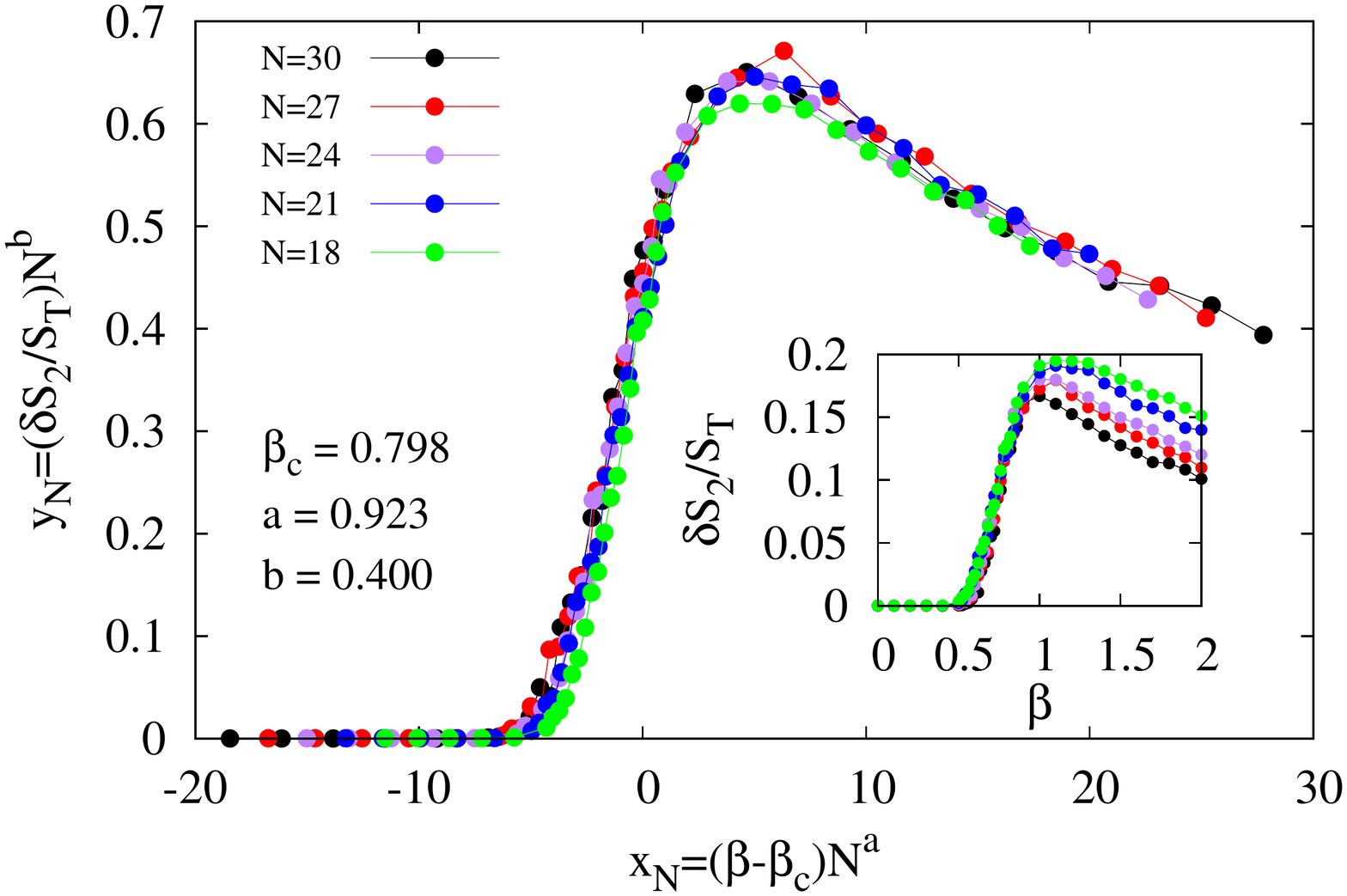}}
\caption{(Color online). a) Scaling collapse of $\langle S_2 \rangle /S_T$ at $t=1/3$, where $S_T=N_A \log2$. It can be noticed that $b$ is very close to 0. b) Scaling collapse of $\delta S_2/S_T$. The left insets show the original curves. Two scaling collapses give very close $\beta_c$ values. Error analysis is only performed on the scaling collapse of $\langle S_2 \rangle /S_T$.}
\label{S2_scaled}
\end{figure}

\begin{figure}[hbt]
\centering
\includegraphics[scale=.32]{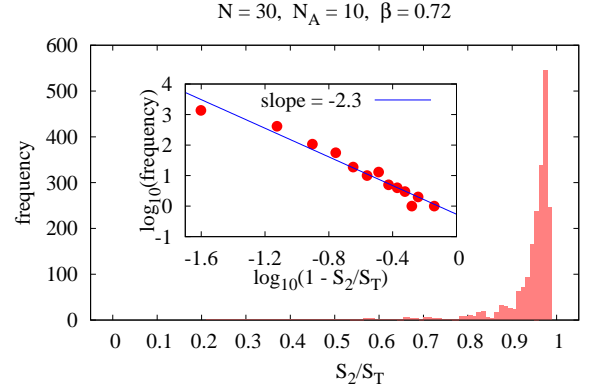}
\caption{(Color online) The distribution of $S_2/S_T$ at $\beta=0.72$. The total number of samples is 2000. The number of bins used to make the histogram is 100. The inset shows a log-log plot with a linear fit, which indicates a power law distribution. The number of bins in the log-log scale is 20 for $S_2/S_T$ from 0 to 1. However, as many of the bins have 0 sample, some points are not included the log-log plot.}
\label{B07fluc}
\end{figure}

To locate the MBL phase transition, we use both $\langle S_2 \rangle$ and its standard deviation, $\delta S_2$.\cite{Kjall2014, Vosk2014} For instance, in the bottom inset of Fig.~\ref{S2_scaled}, it is shown that when $\beta>0.5$, $\delta S_2/S_T$ will increase rapidly and reach the maximum value at some $\beta$.   Fig.~\ref{S2_scaled} (bottom) shows that the standard deviation at intermediate values of $\beta$ is actually a non-trivial fraction of the maximum thermodynamic entropy.   In fact, because of the breakdown in the REM of the Central Limit Theorem, the fluctuation of the R\'enyi entropy is non-Gaussian. This is seen, for example, in the distribution of $S_2/S_T$ at $\beta=0.72$ shown in Fig.\ref{B07fluc}, which  is peaked around 1 and has a long tail. This long tail has power law scaling behavior  shown in the inset of Fig.\ref{B07fluc}, with a power-law exponent between -3 and -2,  which implies a well-defined average but an infinite variance in the thermodynamic limit (which is consistent with the peak scaling as $S_T$).  This may be related to the quantum Griffiths phase found in Ref.~\onlinecite{Vosk2014, Agarwal2015,Potter2015}.

%%%%%%%%%%%%%%%%%%%%
%%%%%%%%%%%%%%%%%%%%
%%%%%%%%%%%%%%%%%%%%
%%%%%%%%%%%%%%%%%%%%

%%%%%%%%%%%%%%%%%%%%
%%%%%%%%%%%%%%%%%%%%

While we can use the peak of $\delta S_2$ and the transition of $\langle S_2\rangle$ from maximal to zero to locate the phase transition, these quantities scale with system size. We therefore use finite-size scaling to find the MBL phase transition. In the regime of interest, we perform a scaling collapse of the data for the two ratios $\langle S_2\rangle /S_T$ and $\delta S_2/S_T$ separately,  using  scaling functions of the  form
\begin{equation}
N^b \Phi((\beta-\beta_c)N^a)
\label{scaling_form}
\end{equation}
and determine the form of the scaling functions $\Phi(x)$ numerically. The exponent $b$ is expected to be very close to 0 for $\langle S_2\rangle/S_T$. When doing the scaling analysis, we find that the quality of the $\langle S_2\rangle /S_T$ collapse is  better than that of the $\delta S_2/S_T$. Error analysis is only performed on the scaling parameters obtained from $\langle S_2\rangle /S_T$, although generally the $\delta S_2/S_T$ collapse yields similar values for $\beta_c$.

Notice that this scaling form assumes the existence of only one transition in spite of the fact that we have analytical bounds that show the presence of ITV, constant, and sub-ITV scaling; numerically the latter appears to be neither constant nor linear with subsystem size.

\begin{figure}[hbt]
\centering
\includegraphics[scale=.32]{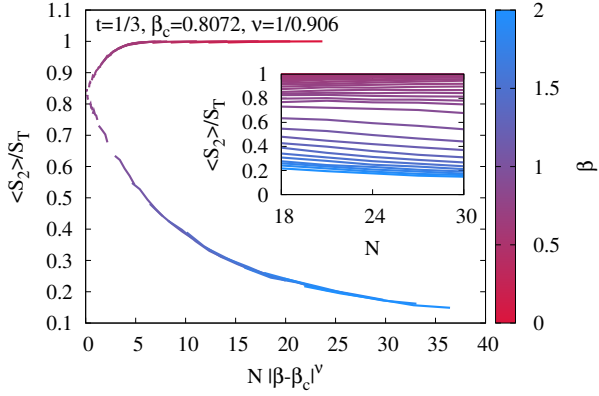}
\caption{(Color online) This graph uses the scaling parameters obtained from Fig. \ref{S2_scaled} to scale the $\langle S_2 \rangle/S_T$ vs $N$ curves for $\beta$ values ranging from 0 to 2. The scaled graph clearly shows two branches -- (1) curves with large $\beta$ flow to low entanglement entropy; (2) curves with small $\beta$ flow to $T=\infty$ entanglement entropy, i.e. $N_A\log2$. The right inset shows the original curves, where each curve corresponds to a different $\beta$. }
\label{DL_S2_scaled}
\end{figure}

Fig.~\ref{S2_scaled} shows the scaling collapsed for $\langle S_2\rangle $ at $t=1/3$.  In previous work on MBL phase transition, $\nu \equiv a^{-1}$ in Eq.\eqref{scaling_form}, is the critical exponent for the localization length and is expected to satisfy Harris inequality $\nu\geq 2/d$, where $d$ is the spatial dimension. However REM model is a highly non-local model and $d$ here is equal to infinity, which suggests that there is no bound for $\nu$. To make sure the scaling parameters obtained are truly associated with this transition, one can use them to scale the $\langle S_n\rangle /S_T$ data following Ref. \onlinecite{Luitz2015}. For each $\beta$, we plot  $\langle S_n\rangle$ vs $N$  as shown in the inset of Fig.~\ref{DL_S2_scaled} , from which one can see that the scaled curves bifurcate smoothly into the two branches depending on $\beta$. Furthermore, Fig.~\ref{DL_S2_scaled} indicate that the system experiences a phase transition rather than a crossover, because of the clear separation of the two branches at large $N$.

%%%%%%%%%%%%%%%%%%%%
%%%%%%%%%%%%%%%%%%%%

%%%%%%%%%%%%%%%%%%%%
\begin{figure}[hbt]
\centering
\subfigure[]{\includegraphics[scale=.3]{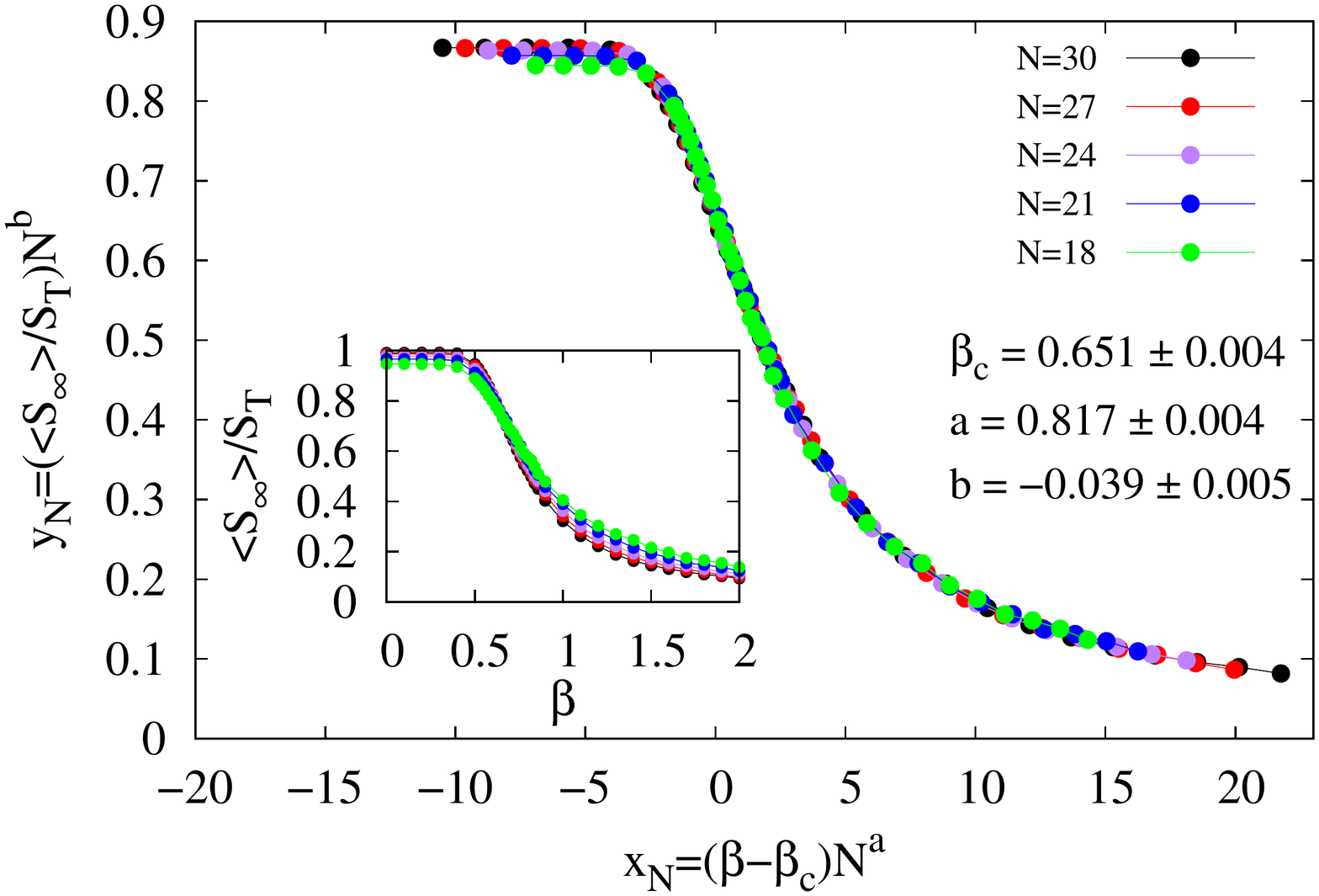}}
\subfigure[]{\includegraphics[scale=.3]{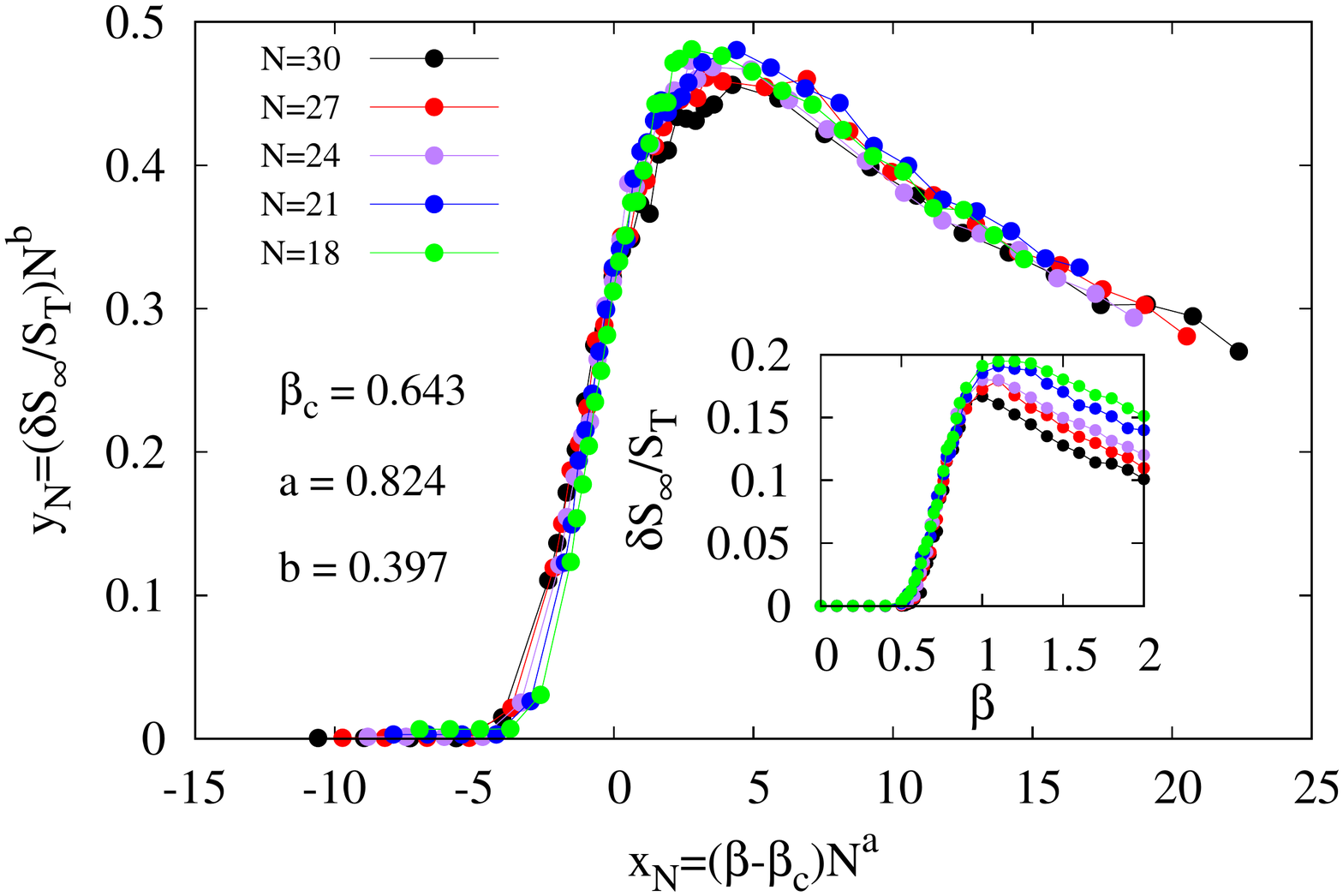}}
\caption{(Color online) (a)  Scaling collapse of $\langle S_{\infty} \rangle /S_T$ at $t=1/3$, where $S_T=N_A\log2$. It can be noticed that $b$ is very close to 0. (b) Scaling collapse of $\delta S_{\infty}/S_T$. The left insets show the original curves. Two scaling collapses give very close $\beta_c$ values. Error analysis is only performed on the scaling collapse of $\langle S_{\infty} \rangle /S_T$.}
\label{Sinf_scaled}
\end{figure}
%%%%%%%%%%%%%%%%%%%%

%%%%%%%%%%%%%%%%%%%%
\begin{figure}[hbt]
\centering
\includegraphics[scale=.32]{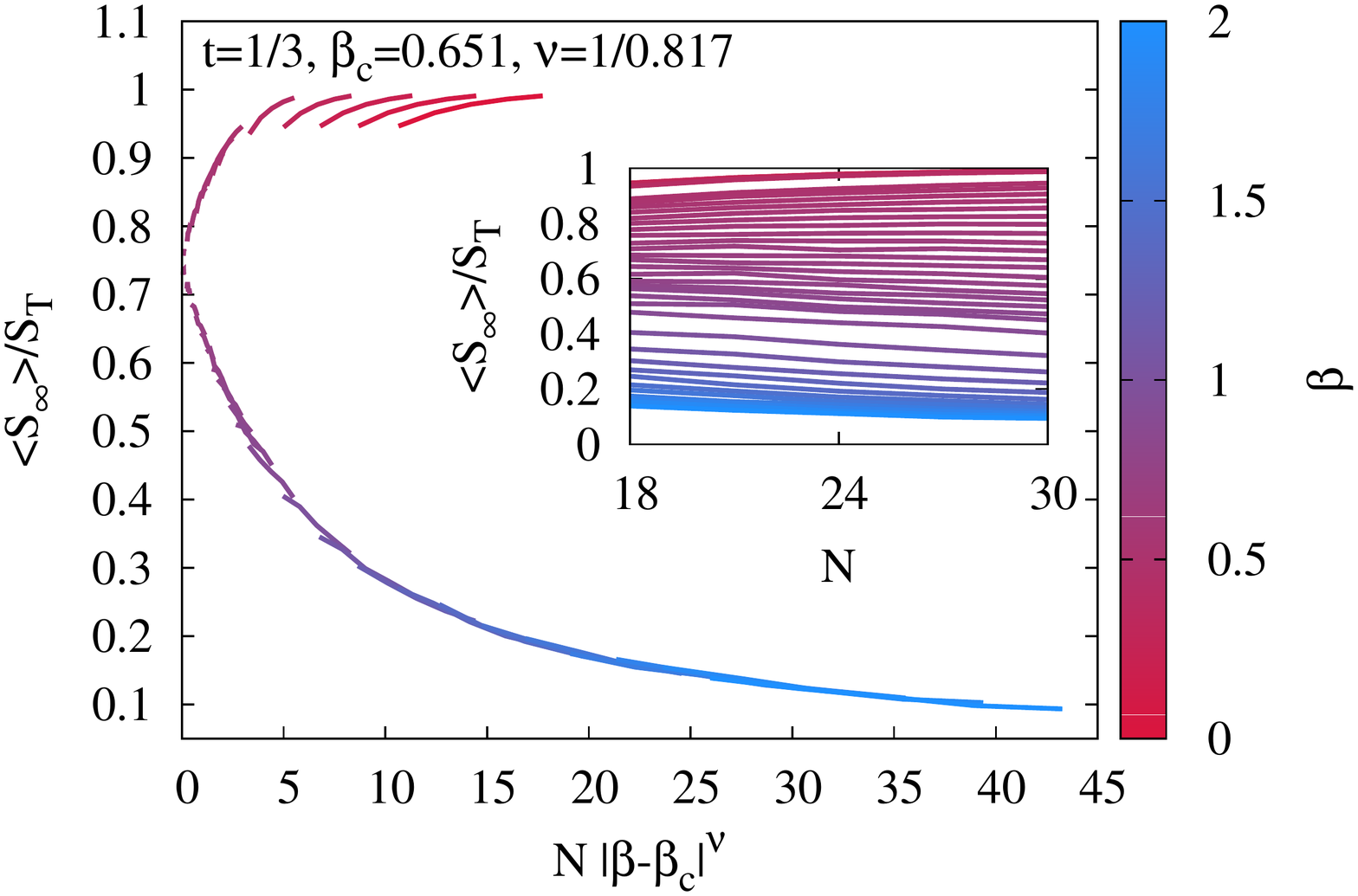}
\caption{(Color online) This graph uses the scaling parameters obtained from Fig. \ref{Sinf_scaled} to scale the $\langle S_{\infty} \rangle/S_T$ vs $N$ curves for $\beta$ values ranging from 0 to 2. The scaled graph clearly shows two branches -- (1) curves with large $\beta$ flow to low entanglement entropy; (2) curves with small $\beta$ flow to $T=\infty$ entanglement entropy, i.e. $N_A \log2$. The right inset shows the original curves, where each curve corresponds to a different $\beta$. }
\label{DL_Sinf_scaled}
\end{figure}
%%%%%%%%%%%%%%%%%%%%

%%%%%%%%%%%%%%%%%%%%
\begin{figure}[hbt]
\centering
\includegraphics[scale=.32]{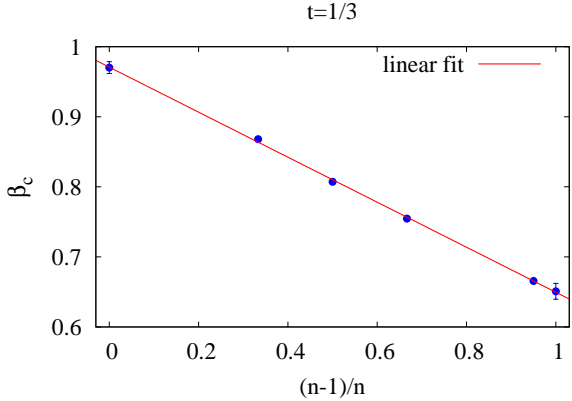}
\caption{ (Color online) $\beta_c$ vs $\frac{n-1}{n}$ curve at $t=1/3$. The linear fit line's slope is $-0.3213 \pm 0.0034$, and its y-intercept is $0.9709 \pm 0.0023$.}
\label{betac_n}
\end{figure}
%%%%%%%%%%%%%%%%%%%%

We have also done scaling collapse for $\langle S_2\rangle$ with other subsystem ratios. At $t=1/2$, we get $\beta_c=0.693$; and at $t=1/5$, we get $\beta_c=0.965$. These results are shown as purple dots on Fig. \ref{summary}. All our collapsing results are summarized in Table \ref{Scaled_Data}.

\begin{table}[htbp]
\centering
\begin{tabular}{|c|c|c|c|c|c|}
\hline
 &$t$ & $n$ & $a$ & $b$ & $\beta_c$ \\
\hline
 & 1/3 & 1 & 0.811 & 0.007 & 0.970 \\ \cline{2-6}
 & 1/3 & 1.5 & 0.897 & 0.0093  & 0.8680 \\ \cline{2-6}
 & 1/3 & 2 & 0.906 & −0.0006  & 0.8072 \\ \cline{2-6}
$\frac{\langle S_n \rangle}{S_T}$ & 1/3 & 3 & 0.902 & 0.0095 & 0.755 \\ \cline{2-6}
 & 1/3 & 20 & 0.8389 & −0.025 & 0.6657 \\ \cline{2-6}
 & 1/3 & $\infty$ & 0.817 & −0.039 & 0.651 \\ \cline{2-6}
 & 1/2 & 2 & 0.90 & −0.070 & 0.693 \\ \cline{2-6}
 & 1/5 & 2 & 0.766 & 0.010 & 0.965 \\ \cline{2-6}
\hline\hline
 & 1/3 & 1 & 0.893 & 0.321 & 0.983 \\ \cline{2-6}
 & 1/3 & 1.5 & 0.902 & 0.396 & 0.861 \\ \cline{2-6}
 & 1/3 & 2 & 0.923 & 0.400 & 0.798 \\ \cline{2-6}
$\frac{\delta S_n}{S_T}$ & 1/3 & 3 & 0.908 & 0.399 & 0.753 \\ \cline{2-6}
 & 1/3 & 20 & 0.835 & 0.383 & 0.655 \\ \cline{2-6}
 & 1/3 & $\infty$ & 0.824 & 0.397 & 0.643 \\ \cline{2-6}
 & 1/2 & 2 & 0.909 & 0.090 & 0.678 \\ \cline{2-6}
 & 1/5 & 2 & 0.856 & 0.147 & 0.963 \\ \cline{2-6}
\hline
\end{tabular}
\caption{Collected scaling collapse data}\label{Scaled_Data}
\end{table}

Based on the above numerical results, we conclude that
% as long as the ratio $t$ takes finite value, 
the MBL transition (both of $\langle S_2\rangle $ and all the other $\langle S_n\rangle$ (see Sec.\ref{sec:multi_Renyi} and Appendix C) happen separately from the localization transition at $\beta_{sg}$.  The MBL phase transition is at $\beta_{c}$ and is smaller than $\beta_{sg}$. Between $\beta_c$ and $\beta_{sg}$, although it is in the MBL phase, the whole wave function is still delocalized in the Hilbert space. Only when $\beta>\beta_{sg}$, the wave function becomes localized in the Hilbert space and is already deep in the MBL phase.

%\textcolor{blue}{EF: these are strong statements and we need to provide evidence for it here.}

\subsection{Intermediate regime}
\label{sec:intermediate}

According to the scaling collapse in Fig. \ref{S2_scaled}, the MBL phase transition for $\langle S_2\rangle$ happens at around $0.8$ at $t=1/3$. From Eq.\eqref{upper_b}, we also know that when $\beta>0.697$, $\langle S_2\rangle<N_A\log2$. This implies that there is an intermediate regime  between ITV and MBL phase (regime (ii) in Fig.\ref{fig:schematic} (b)). In this regime, $\langle S_2\rangle$ is sub-extensive and the slope
\begin{align}
0<s(N_A)=\frac{1}{\log2}\frac{d\langle S_2\rangle}{d N_A}<1.
\end{align}
This regime is not a cross-over but sharply defined with non-analyticities in the curve $s(t)$ signaling its
beginning. While a combination of our analytical bounds and numerical results can bound the location of this transition, we are not able to numerically pinpoint its location.  It is interesting to note, though, 
that all the curves in Fig.~\ref{S2_slope} seem to cross at a single point;  for $t=1/3$, this point is approximately
 $\beta=0.72$ which is surprisingly close to the $\beta_2$ bound.
While we can't say anything definitive about the value of the $s(N_A)$ in the intermediate regime, Fig.~\ref{S2_slope} shows only two plateaus suggesting that the $s(N_A)$ in this intermediate regime is not
constant.  Instead we conjecture that the slope changes continuously as a function of $N_A$ in this regime in a nonlinear way.  This means that this regime is non-thermal and doesn't correspond to a thermal density matrix at any temperature. It is not clear whether to call this regime a separate phase, particularly as the data collapse
on $\langle S_2\rangle / S_T$ and $\delta S_2/S_T$ only identify a single transition.

\subsection{Multifractality of the R\'enyi entanglement entropies}
\label{sec:multi_Renyi}

While we have identified transitions in $\langle S_2 \rangle$, we can also consider $\langle S_n \rangle$ for $n \neq 2$. 
%The scaling behavior of different $\langle S_n\rangle$ in the intermediate regime is complicated.  
The analytical bounds (Fig.\ref{S_n_bound}) show that there is a regime where $S_2$ is still ITV where
$S_\infty$ scales at a rate less then ITV.  We can also use scaling collapse to identify the MBL phase
transition in $\langle S_n\rangle$;  for example, see  Fig.~\ref{Sinf_scaled} and \ref{DL_Sinf_scaled}.  More scaling collapse graphs with different $n$ and subsystem ratio $t$ can be found in Appendix C.  We summarize the $N_A/N=1/3$ results for various $n$ in Fig.\ref{betac_n}, from which one can clearly see the transition depends on $n$.  We attribute this as a sign  of multifractal behavior similar to the multifractal behavior found in the critical wave function of the Anderson localization phase transition point. \cite{Wegner1980} 
In the Anderson localization problem it is known that multifractality is a feature of the wave function for the mobility edge.\cite{Wegner1980} It is not known if this is also the case in  MBL or if there is a multifractal phase. From our data we cannot at present make a definitive determination. It is interesting to note that $\beta_c(n)$ is linearly proportional to $(n-1)/n.$  The phase transition point $\beta_c$ for von Neumann entropy is $0.97$ and we identify this as the true MBL transition. 

\begin{figure}[hbt]
\centering
\includegraphics[scale=.32]{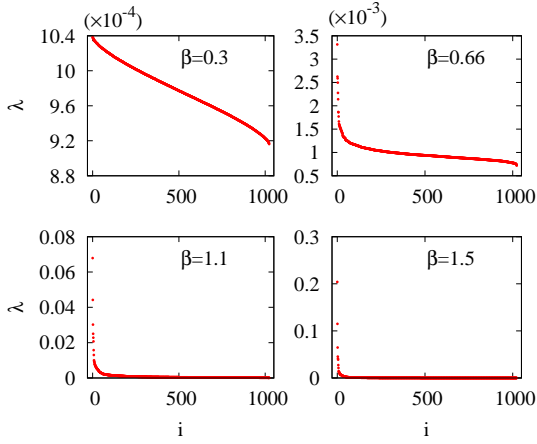}
\caption{(Color online) Entanglement spectrum for one randomly chosen disorder configuration at $\beta=0.3, 0.66, 1.1$ and $1.5$, with $N=30$, $N_A=10$. The gaps in the entanglement spectrum become evident when $\beta$ gets larger.}
\label{E_gap}
\end{figure}

We can understand the different scaling behavior of $\langle S_n\rangle$  by considering the entanglement spectrum. Fig.\ref{E_gap} is the distribution of eigenvalue $\lambda$ of $\rho_A$ for a randomly chosen disorder configuration at different $\beta$. When $\beta=0.3$, $\lambda$ forms a continuous band around $0.001$ which is approximately equal to $1/2^{N_A}$. While for $\beta=0.66,1.1$ and $1.5$, there is an obvious gap between the lower continuous band and the other higher eigenvalues.  The entanglement gap increases as $\beta$ increases.

This inspires us to write down a simplified two-level model for entanglement spectrum which includes a flat band and a single $\lambda_{max}$. If we assume that $\langle S_n\rangle$ is approximately equal to its upper bound in Eq.\eqref{S_n_upper}, by using $S_{n=\infty}=-\log\lambda_{max}$, we have 
\begin{equation}
\lambda_{max}=2^{-(1-\frac{\beta}{\sqrt{2\log 2}})^2N}
\end{equation}
All the other $\lambda_1\approx 1/2^{N_A}$ and there is a gap between $\lambda_1$ and $\lambda_{max}$. This toy model exhibits the multifractal behavior described above with a transition to constant entanglement slope that scales with $n>2$. The transition in this simplified model is direct from ITV to constant;  to capture our intermediate regime, the single $\lambda_{max}$ can be replaced by a finite number $\lambda_i$ with each of them are separated by a finite gap. Also the lower flat band can be replaced a continuous band with more complicated band structure. These additional ingredients are required to have an intermediate regime with sub ITV as well as accurately finding the constant $n=1$ von Neumann entropy.

%%%%%%%%%%%%%%%%%%%%
%%%%%%%%%%%%%%%%%%%%

\subsection{The random sign structure in the wave function}
\label{sec:random-sign-structure}

\begin{figure}[hbt]
\centering
\includegraphics[scale=.32]{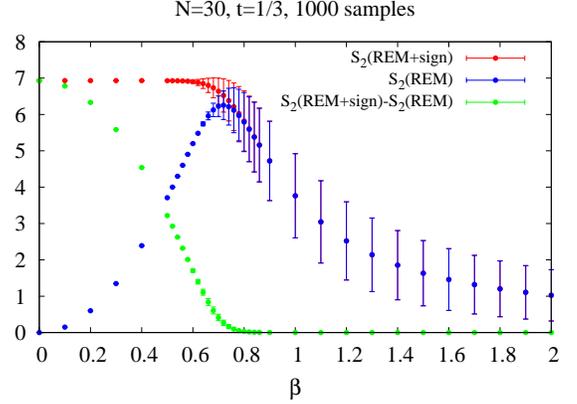}
\caption{(Color online) $\langle S_2\rangle$ of REM+\textrm{sign} and REM, and their difference. The 1000 disorder configurations are the same for REM+\textrm{sign} and REM. Error bars here represent standard deviations. The difference of the $\langle S_2\rangle$ has well bounded variance for all $\beta$.}
\label{S_compare}
\end{figure}

We have introduced a random sign structure to convert our ground state wave function into one at finite energy density.  While for any strictly positive wave function, the introduction of random signs can only increase the entanglement entropy, it is interesting to ask what effect, if any, the random sign has here.  This requires computing the R\'enyi entropy $\langle S_n(\rho_A^{\prime})\rangle$ for the REM wave function without the random sign $|\Psi_{REM}\rangle$. Different from $|\Psi_{REM+\textrm{sign}}\rangle$, the R\'enyi entropy for $|\Psi_{REM}\rangle$ is not a monotonic function with $\beta$ and has more complicated scaling behavior.  In fact $|\Psi_{REM}\rangle$ has zero entanglement entropy at both $\beta=0$ and $\beta \rightarrow \infty$.   
It is interesting to note that while  $\langle S_2(\rho_A)\rangle$ and $\langle S_2(\rho_A^{\prime})\rangle$ have long tails and infinite variance 
in the thermodynamic limit at intermediate $\beta$, their difference $\Delta S_2= \langle S_2(\rho_A)\rangle - \langle S_2(\rho^{\prime}_A) \rangle$ appears to have finite variance (as shown in  Fig. \ref{S_compare}).

%%%%%%%%%%%%%%%%%%%%
%%%%%%%%%%%%%%%%%%%%

Since the quenched average $\langle S_2(\rho_A^{\prime})\rangle$ is hard to access analytically, we can compute a lower bound via  Eq.\eqref{anneal} using Jensen's inequality\cite{Reed-Simon}
(see Appendix B for details).  We find that if $t<1/3$, there is a region $\sqrt{2t \log 2} \leq \beta \leq \sqrt{(1-t)\log 2}$ where $|\Psi_{REM}\rangle$ has ITV entanglement. As this is the maximal allowed value, it then directly follows that there is no difference between $\langle S^{\prime}_2\rangle$ and $\langle S_2\rangle$ due to the introduction of signs. 

\begin{figure}[hbt]
\centering
\includegraphics[scale=.32]{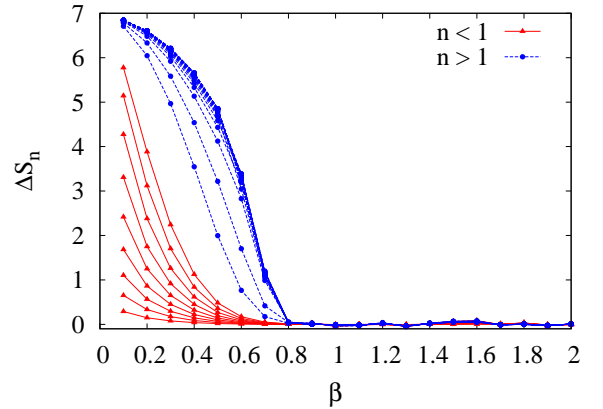}
\caption{(Color online) Difference of $\langle S_n\rangle$ between REM+\textrm{sign} and REM for  $n=0.1,\ldots,0.9$ in $0.1$ steps (triangles), and for $n=1.5$, $n=2, \ldots,20$, and $\infty$ (full circles), at $N=30$ and $N_A=10$. The number of disorder configurations is 2000. The random configurations are not the same for REM+\textrm{sign} and REM.}
\label{diff}
\end{figure}

\begin{figure}[hbt]
\centering
\includegraphics[scale=.32]{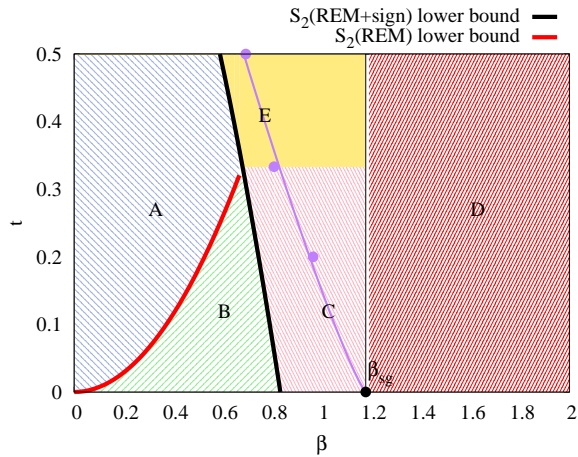}
\caption{(Color online) Different behaviors of $\Delta S_2 = \langle S_2(REM+\textrm{sign}) \rangle - \langle S_2(REM) \rangle$ on the $t$ vs $\beta$ graph.
Region A represents the area where $\langle S_2(REM+\textrm{sign}) \rangle$ follows ITV and $\langle S_2(REM) \rangle$ is a constant, so $\Delta S_2$ obeys ITV.
Region B represents the area where $\langle S_2(REM+\textrm{sign}) \rangle$ and $\langle S_2(REM) \rangle$ are both ITV, and $\Delta S_2$ is zero. As a result,  $\Delta S_2$ is also zero in Region C based on the monotonicity argument. 
In Region D, both $\langle S_2(REM+\textrm{sign}) \rangle$ and $\langle S_2(REM) \rangle$ are localized in Hilbert space and $\Delta S_2$ is zero. We do not have enough information to determine the behavior $\Delta S_2$ in Region E.}
\label{fig:signDiff}
\end{figure}

Moreover, we conjecture that $\Delta S_2$ decreases monotonically as a function of $\beta$.  This is consistent with the numerical results shown in Fig.\ref{diff}.  Following from this conjecture, we would have that $\Delta S_2=0$ for all $\beta\geq\sqrt{2t\log2}$ and $t<1/3$.  This is because, for all $t<1/3$ both models show ITV between $\beta=\sqrt{2t\log2}$ and $\sqrt{(1-t)\log 2}$ and hence $\Delta S_2=0$.  The regions B, C, D in Fig.~\ref{fig:signDiff} denote  where  $|\Psi_{REM}\rangle$ and $|\Psi_{REM+\textrm{sign}}\rangle$ have the same $\langle S_2\rangle$.

%%%%%%%%%%%%%%%%%%%%
%%%%%%%%%%%%%%%%%%%%

%%%%%%%%%%%%%%%%%%%%
%%%%%%%%%%%%%%%%%%%%

\begin{figure}[hbt]
\centering
\includegraphics[scale=.32]{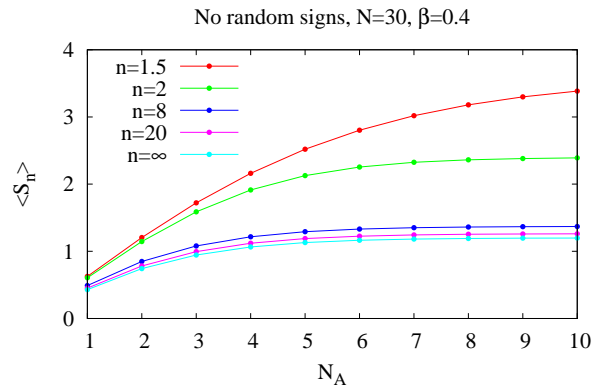}
\caption{(Color online) The numerical results for $\langle S_n^{\prime}\rangle$ without random sign with $n$ from $1.5$ to $\infty$ at $\beta=0.4$. The total system size is $N=30$. Each point in the plot is the average of 2000 disorder realizations.}
\label{NoSg_B04}
\end{figure}

Having identified regimes where the introduction of random signs doesn't affect the entanglement entropy, we also identify regimes where the entanglement entropy can  be shown to be different.  When $\beta=0$, $|\Psi_{REM}\rangle$ is a constant (actually a product state) whereas $|\Psi_{REM+\textrm{sign}}\rangle$ is a volume law.   We can argue that this extends to larger $\beta$.  Defining $X=\mbox{Tr}\widetilde{\rho}_A^2$ and $Y=\mbox{Tr}(\widetilde{\rho}_A^{\prime})^2$, where $\widetilde{\rho}_A$ is the unnormalized reduced density matrix, we have
\begin{equation}
\langle \log\frac{X}{Y}\rangle\leq\log\langle\frac{X}{Y}\rangle\leq\log\frac{\langle X\rangle}{\langle Y\rangle}
\end{equation}
where the second inequality, while not true in general, appears to be numerical validated in our case. In the thermodynamic limit, $\log\frac{\langle X\rangle}{\langle Y\rangle}$ can be directly computed (see Eq.\eqref{random_sign} and Eq.\eqref{no_random_sign}).   We find that when $t<1/3$ and $\beta\leq\sqrt{2t\log2}$,  $\log\frac{\langle X\rangle}{\langle Y\rangle}=0$.  In this region, $\langle S_2\rangle$ for $|\Psi_{REM+\textrm{sign}}\rangle$  continues growing as volume  while $|\Psi_{REM}\rangle$ stays constant,  indicating that the random sign structure can thermalize the wave function and is responsible for the volume law scaling behavior. Fig.~\ref{NoSg_B04} is the numerical result for $\langle S_n^{\prime}\rangle$ at $\beta=0.4$. We can see that when $n\geq 2$, they all saturate to a constant. This region is highlighted in blue on  Fig.~\ref{fig:signDiff} and marks a region where the models differ maximally. 
% Moreover, if $|\Psi_{REM}\rangle$ is a constant at one $N_A$ it should continue to be a constant for larger $N_A$ (\textcolor{red}{For $n=1$ case, we can use strong subadditivity to argue it}).  
%This region is highlighted in  yellow on Fig.~\ref{fig:signDiff}.

Finally, we note that $\Delta S_n$ can be  numerically computed.  Fig.\ref{diff} is $ \Delta S_n$ for $N=30$ and $N_A=10$. We find that for all $n$,  $\Delta S_n=0$ when $\beta>0.8$. This result, while only for $N=30$ and so not absent finite-size effects happens to be at the location of the $\langle S_2\rangle$ MBL phase transition.

%%%%%%%%%%%%%%%%%%%%
%%%%%%%%%%%%%%%%%%%%

\section{Conclusions}
\label{sec:conclusions}

We have studied the many-body localization phase transition in a class of many-body wave functions. We focused our analysis in a class of wave functions,  $|\Psi_{REM}\rangle$, whose amplitudes are the Boltzmann weights of a classical spin glass model in infinite dimension, the Random Energy Model. In order to mimic the structure of wave functions of highly excited states with a finite excitation energy density we considered  another class of states,  $|\Psi_{REM+\textrm{signs}}\rangle$, whose amplitudes are obtained by multiplying the amplitudes of $|\Psi_{REM}\rangle$ by a random sign for each configuration. We studied the MBL problem in the   $|\Psi_{REM+\textrm{signs}}\rangle$ wave function, for different regime of the parameter $\beta$,
 by using both analytical and numerical approaches. 
 
  We showed that there is a direct phase transition into the MBL phase. Here we assume that  the MBL phase is characterized that the entanglement entropies, as a function of the size of the observed region $N_A$, scale  to a constant value, a feature that we observed explicitly for large enough values of $\beta$.
   The location of the phase transition point is identified by scaling collapse of R\'enyi entropy and its standard deviation. In the thermalized regime, there is a regime where the R\'enyi entropies with different R\'enyi index all equal to the thermal entropy $S_T$ at $T=\infty$. When $\beta>\beta_c$, the system enters into the MBL phase where the entanglement entropy is sub-extensive. The MBL phase transition point $\beta_c$ is smaller than $\beta_{sg}$ and this suggests that the MBL phase transition and the classical spin glass phase transition are  different. Upon entering  the MBL phase, the random sign structure is not important any more and $\langle\Delta S_n\rangle$ between $|\Psi_{REM+\textrm{sign}}\rangle$ and $|\Psi_{REM}\rangle$ is zero. For $\beta>\beta_{sg}$, the wave function is deep inside the MBL phase and only $O(1)$ number of the configurations in the REM wave function contributes significantly to the statistical average. The R\'enyi entropy in this regime will reduce to a finite constant.
We find that close to the phase transition point $\beta_c$, the fluctuation of R\'enyi entropy is strong in the finite size system. In this regime, R\'enyi entropies with different R\'enyi index show different scaling behavior and are similar to the multifractal behavior observed at the Anderson localization phase transition point. Moreover, $\langle S_n\rangle$ has a phase transition at different $\beta_c$.
Finally we note that we have refrained ourselves from performing the same extensive studies for  the wave function without random signs, $|\Psi_{REM}\rangle$. While we have evidence that this wave function too has a thermalized regime, since it has only strictly positive amplitudes we do not expect it to provide an useful description of the MBL problem. Nevertheless it may be useful to investigate its properties in a separate publication. % and we identify $\beta_c$ for von Neumann entropy as the true MBL phase transition point.

\textit{Note added in the proof}: After this work was accepted for publication we became aware of the recent independent work of Yang, Chamon, Hamma and Mucciolo\cite{Yang-2015} who have also discussed the spectrum of the reduced density matrix in several models, including the wave functions discussed in this work.

\begin{acknowledgments}
We thank Dmitry Abanin, David Huse, and Shivaji Sondhi for useful conversations and comments. EF thanks the KITP (and the Simons Foundation) and its  ENTANGLED15 program for support and hospitality. This work was supported in part by the National Science Foundation grants DMR-1064319 and DMR-1408713 (XC,GYC,EF)  at  the University of Illinois, PHY11-25915 at KITP (EF),   DOE, SciDAC FG02-12ER46875 (BKC and XY), and the Brain Korea 21 PLUS Project of Korea Government (GYC).  This research is part of the Blue Waters sustained-petascale computing project, which is supported by the National Science Foundation (award numbers OCI-0725070 and ACI-1238993) and the state of Illinois. Blue Waters is a joint effort of the University of Illinois at Urbana-Champaign and its National Center for Supercomputing Applications.  
\end{acknowledgments}
 
\appendix

\section{Entanglement entropy for the RK-wave function with classical local Hamiltonian}
\label{app:EE-RK}

For the RK-wave function defined in Eq.\eqref{wave1}, if the related classical model has a {\em local}  Hamiltonian, the entanglement entropy satisfies the area law. This scaling behavior only relies on the property of the RK state and does not depend on whether the quantum model is critical or not. Here, we will briefly review the calculation following Ref.\onlinecite{Stephan-2012}. 

For the RK state with a classical local Hamiltonian, after partitioning the system into two parts A and B, it can be approximately written in this way,
\begin{equation}
|\Psi_0\rangle=\sum_{\Gamma}\lambda_{\Gamma}|\Psi_A(\Gamma)\rangle|\Psi_B(\Gamma)\rangle
\label{schm}
\end{equation}
where $\Psi_A(\Gamma)$ and $\Psi_B(\Gamma)$ are the RK wave functions defined in region A and region B with the same boundary configuration $\Gamma$
\begin{eqnarray}
\nonumber &&|\Psi_A(\Gamma)\rangle=\sum_{A}\frac{e^{-\frac{\beta}{2}H_A(\Gamma)}}{\sqrt{\mathcal{Z}_A(\Gamma)}}|c_A(\Gamma)\rangle\\
&&|\Psi_B(\Gamma)\rangle=\sum_{B}\frac{e^{-\frac{\beta}{2}H_B(\Gamma)}}{\sqrt{\mathcal{Z}_B(\Gamma)}}|c_B(\Gamma)\rangle
\end{eqnarray}
When $\Gamma\neq \Gamma^{\prime}$, they satisfy $\langle\Psi_A(\Gamma)|\Psi_A(\Gamma^{\prime})\rangle=0$ and $\langle\Psi_B(\Gamma)|\Psi_B(\Gamma^{\prime})\rangle=0$. The summation in Eq.\eqref{schm} is the sum over all possible boundary configurations along the cut and $\lambda_{\Gamma}=\sqrt{\mathcal{Z}_A(\Gamma)}\sqrt{\mathcal{Z}_B(\Gamma)}/\sqrt{\mathcal{Z}}$. 

 Thus Eq. \eqref{schm} is the Schmidt decomposition of the wave function, the reduced density matrix in regime A is
\begin{equation}
\rho_A=\sum_{\Gamma}\lambda_{\Gamma}^2|\Psi_A(\Gamma)\rangle\langle \Psi_A(\Gamma)|
\end{equation}
Since the dimension of $\rho_A$ only depends on the dimension of the Hilbert space along the boundary, the entanglement entropy should satisfy the area law.

\section{Lower bound for $\langle S_2(\rho_A^{\prime})\rangle$}
\label{sec:lower-bound}

For the REM wave function without random sign, the quenched average $\langle S_2(\rho_A^{\prime})\rangle$ is hard to access analytically, instead we calculate the annealed average $S_2(\langle \rho_A^{\prime}\rangle)$ defined in Eq.\eqref{anneal}, which gives the lower bound for $\langle S^{\prime}_2\rangle$. To obtain $S_2(\langle \rho_A^{\prime}\rangle)$, we need to know $\langle\mbox{Tr}(\widetilde{\rho}_A^{\prime})^2\rangle$ first, where $\widetilde{\rho}_A^{\prime}$ is the unnormalized reduced density matrix.
\begin{align}
 \langle\mbox{Tr}(\widetilde{\rho}_A^{\prime})^2\rangle=&2^Ne^{2N\beta^2}
 +(2^{N_B}-1)2^{N}e^{N\beta^2}
 \nonumber\\
 +(2^{N_A}-1)&2^{N}e^{N\beta^2}
+(2^{N_A}-1)(2^{N_B}-1)2^Ne^{\frac{N\beta^2}{2}}
\label{no_random_sign}
\end{align}
This gives $S_2\langle(\rho_A^{\prime})\rangle$ in the thermodynamic limit. When $0<t\leq 1/3$, there are three regimes,
\begin{equation}
S_2(\langle {\rho}_A^{\prime}\rangle)=\begin{cases} \frac{\beta^2N}{2}, & \beta\leq\sqrt{2t\log2}\\
N_A\log2, &\sqrt{2t\log2}<\beta\leq \sqrt{(1-t)\log2}\\
N(\log2-\beta^2), & \beta>\sqrt{(1-t)\log2}
\end{cases}
\end{equation}

When $1/3<t<1/2$, there are two regimes
\begin{equation}
S_2(\langle {\rho}_A^{\prime}\rangle)=\begin{cases} \frac{\beta^2N}{2}, & \beta\leq\sqrt{2\log2/3}\\
N(\log2-\beta^2), & \beta>\sqrt{2\log2/3}
\end{cases}
\end{equation}


%\bibliography{biblio}

\end{document}